\documentclass{llncs}
\usepackage{cite}
\usepackage{graphicx}
\usepackage{textcomp}
\usepackage{xcolor}
\usepackage{graphicx}
\usepackage{caption}
\usepackage{subcaption}
\usepackage{hyperref}
\usepackage{amsmath,amssymb,amsfonts,bm}
\usepackage{stmaryrd}
\usepackage{multirow}
\usepackage{algorithm}
\usepackage[noend]{algpseudocode}
\usepackage{ulem}
\usepackage{tikz}
\usepackage{booktabs}
\usepackage{makecell}
\usepackage{varwidth}
\usepackage{fullpage}
\usepackage[flushleft]{threeparttable}
\def\BibTeX{{\rm B\kern-.05em{\sc i\kern-.025em b}\kern-.08em
    T\kern-.1667em\lower.7ex\hbox{E}\kern-.125emX}}

\algnewcommand\algorithmicinput{\textbf{Procedure}}
\algnewcommand\INPUT{\item[\algorithmicinput]}

\newcommand{\dd}{Dilithium}
\newcommand{\cc}{CRYSTALS}

\begin{document}
\title{High-Throughput GPU Implementation of \dd\ Post-Quantum Digital Signature}
%
%
\author{Shiyu Shen\inst{1,4} \and
Hao Yang\inst{2} \and
Wangchen Dai\inst{3} \and
Hong Zhang\inst{1,4} \and
Zhe Liu\inst{3,2} \and
Yunlei Zhao\inst{1,4}
}
\authorrunning{S. Shen et al.}
%
\institute{
School of Computer Science, Fudan University, Shanghai, China
\and
College of Computer Science and Technology, Nanjing University of Aeronautics and Astronautics, Nanjing, China
\and
Research Institute of Basic Theories, Zhejiang lab, Hangzhou, China
\and
State Key Laboratory of Cryptology, Beijing
}
\maketitle              
\begin{abstract}
Digital signatures are fundamental building blocks in various protocols to provide integrity and authenticity. 
The development of the quantum computing has raised concerns about the security guarantees afforded by classical signature schemes.
\cc-\dd\ is an efficient post-quantum digital signature scheme based on lattice cryptography and has been selected as the primary algorithm for standardization by the National Institute of Standards and Technology.
In this work, we present a high-throughput GPU implementation of \dd.
For individual operations, we employ a range of computational and memory optimizations to overcome sequential constraints, reduce memory usage and IO latency, address bank conflicts, and mitigate pipeline stalls. 
This results in high and balanced compute throughput and memory throughput for each operation. 
In terms of concurrent task processing, we leverage task-level batching to fully utilize parallelism and implement a memory pool mechanism for rapid memory access.
Considering the impact of varying repetition numbers in \dd\ on the overall execution time and hardware utilization, we propose a dynamic task scheduling mechanism to improve multiprocessor occupancy and significantly reduce execution time. 
Furthermore, we apply asynchronous computing and launch multiple streams to hide data transfer latencies and maximize the computing capabilities of both CPU and GPU.
For \dd2, our implementation demonstrates throughputs of 575k and 1409k operations per second for signing and verification procedures, respectively, on an A100 GPU, corresponding to $111\times$ and $69\times$ speedups over a single-thread CPU implementation.
Across all three security levels, our GPU implementation can concurrently compute ten thousand tasks in less than 32 miliseconds for signing and 15 miliseconds for verification on both commercial and server-grade GPUs. 
This achieves microsecond-level amortized execution times for each task, offering a high-throughput and quantum-resistant solution suitable for a wide array of applications in real systems.

\end{abstract}

\section{Introduction} \label{sec:intro}

Digital signature is a cryptographic primitive that ensures message integrity and authenticity.
As a crucial component of information security, digital signature algorithms are widely adopted in various protocols and applications, including Transport Layer Security (TLS) and blockchain systems.
However, the emergence of quantum computing poses a significant threat to classical signature algorithms like RSA and ECDSA \cite{DBLP:journals/siamcomp/Shor97}.
The security of these algorithms relies on large integer factorization and discrete logarithm problems, which are believed to be vulnerable to quantum attacks. 
While it is uncertain whether a powerful enough quantum computer will be developed within decades, exploring post-quantum digital signatures and their practical applications for long-term security is essential.

In 2016, the National Institute of Standards and Technology (NIST) launched the post-quantum cryptography standardization project to standardize post-quantum digital signature algorithms and public-key encryption/key encapsulation mechanisms (KEM) \cite{NIST.IR.8413}.
Four algorithms, including one KEM and three signature schemes, were finally selected after three rounds of evaluation in 2022, where \cc-\dd\ \cite{DBLP:journals/tches/DucasKLLSSS18,nist-dili} is one of them.
Due to its solid security guarantees and efficiency among all three schemes, NIST recommends \dd\ as the primary choice. 
Many real-world applications are considering deploying \dd\ for long-term protection. 
For example, ArielCoin \cite{arl}, an experimental digital currency, uses \dd\ for verification and authentication; the $\mathsf{liboqs}$ library \cite{liboqs} presented by the Open Quantum Safe project integrates \dd\ for a quantum-safe TLS protocol. 
However, lattice-based schemes like \dd\ suffer from high computational and memory overhead, as well as large IO transfer sizes, which make them a performance bottleneck in server-based scenarios with vast numbers of queries.  
Therefore, there is a need for highly optimized implementations of \dd\ with high throughput and efficient memory management specifically designed for server-based solutions.

GPUs are commonly used for concurrent processing of signatures due to their massive parallelism. 
Many studies have been conducted on GPU-based implementations of classical signatures and post-quantum KEMs, which have shown high performance \cite{DBLP:journals/tpds/GuptaJCC21,DBLP:conf/iscas/OnoBS21,DBLP:journals/tpds/GaoXW22,DBLP:journals/access/LeeSZH22,DBLP:conf/esorics/WanZFWGWLD22}. 
However, there is a lack of research focusing on post-quantum signatures \cite{DBLP:journals/tpds/SunZM20,seo2022parallel}. 
Previous works have used either a single thread or partial threads in a warp to execute tasks, which is not suitable for hiding IO latency as data accesses may not be coalesced. 
This is not suitable for hiding IO latency, as data accesses may not be coalesced.
The implementations of \dd\ on software \cite{DBLP:journals/tches/DucasKLLSSS18,DBLP:conf/cardis/RaviGCB19,DBLP:journals/tches/GreconiciKS21,DBLP:journals/tches/BeckerHKYY22,DBLP:conf/acns/AbdulrahmanHKS22} and hardware \cite{DBLP:journals/tches/ZhaoZWYZLZYWL22,DBLP:journals/tches/BanerjeeUC19,DBLP:journals/iacr/BeckwithNG22} platforms have been studied, but they focus on compact design and are not sufficient to meet the throughput and real-time requirements.

\paragraph{Contributions.}
In this work, we present a high-throughput GPU implementation of \dd.
Our main contributions can be summarized as follows:
\begin{itemize}
	\item For all operations, we develop optimized implementations to enhance performance. 
		Specifically, we parallelize numerous sequential operations in rejection sampling by leveraging CUDA integer intrinsics and warp-level primitives, optimize memory access patterns in number-theoretic transforms, and minimize resource usage in hash functions and inner-product calculations. 
		Furthermore, we implement a memory pool mechanism to enable efficient memory management.
	\item We introduce several optimizations to minimize IO latency and improve resource utilization based on the profiling results of our implementation.
		First, we propose a finely-tuned fusing strategy that strikes a balance between low IO latency and high resource utilization, resulting in optimal performance. 
		Second, we introduce a dynamic task scheduling mechanism to address the occupancy decrease issue when batching multiple signing procedures, which significantly improves resource utilization and reduces execution time.
		Besides, we asynchronize the computation and launch multiple CUDA streams to hide the latency of data transfers between the CPU and GPU, fully utilizing the computational capabilities of both CPU and GPU.
	\item We deploy our implementation on three representative GPU devices with varying computing capabilities. For the signing procedure, we achieve 316k to 575k operations per second throughput on a server-grade NVIDIA Tesla A100 GPU, and 272k to 563k operations per second throughput on a desktop NVIDIA RTX 3090 Ti GPU, which shows at least a $102\times$ speedup compared to single-thread CPU implementation.
\end{itemize}

\paragraph{Code.}
The code of our implementation has been submitted in attachments.
We prepare to open source the whole project after the paper is accepted.

\section{Preliminaries} \label{sec:pre}

\subsection{Notation} \label{sec:pre-notation}
Let $\mathbb{Z}$ be the group of integers.
We define $\mathbb{Z}_q$ with its representation in the interval $\mathbb{Z}\cap [-\frac{q}{2}, \frac{q}{2})$, where $q$ is a prime.
Let $n$ be a power of 2, we use $R = \mathbb{Z}[X]/(X^n+1)$ to denote the $2n$-th cyclotomic ring and $R_q = R/qR$ to denote its residue ring modulo $q$.
We use lowercase letters to represent ring elements (polynomials), e.g., $\bm{f} = \sum_{i=0}^{n-1} f_i X^i$, and vectors, e.g., $\mathbf{v}$. 
Bold uppercase letters represent matrices of polynomials, e.g., $\mathbf{A}$.
We use a hat symbol to indicate elements in frequency domain.

The notation $\bmod^{\pm}\alpha$ denotes centered reduction modulo $\alpha$, which outputs representatives in the range $(-\lfloor\frac{\alpha+1}{2}\rfloor, \lfloor\frac{\alpha}{2}\rfloor]$.
The operator $\|\cdot\|_\infty$ represents the $\ell_\infty$-norm, and $[ \cdot ]$ records coefficient numbers that equal 1. 
The notation $\lfloor \cdot \rfloor$ refers to flooring, and modular reduction by $q$ is denoted by $[\cdot]_q$.
The set notation $\{0, 1\}^l$ indicates an $l$-bit stream; similarly $\{0, 1\}^*$ represents arbitrary length bit streams with $|\cdot|$ indicating their bit-length. 
The operator $\|$, when applied on elements implicitly converts them into bit streams and then concatenates them.
For a finite set $S$, $a \leftarrow S$ refers to uniform sampling from $S$. 
We define $S_\eta := \{\omega: \omega\in R, \|\omega\|_\infty\leq\eta\}$ and $\tilde{S}_\eta := \{\omega\bmod^{\pm} 2\eta:\omega\}$.
The operator $\llbracket \mathcal{P} \rrbracket$, where $\mathcal{P}$ is a statement, returns 1 if $\mathcal{P}$ is true and 0 otherwise.

\subsection{Ring Arithmetic} \label{sec:pre-poly}

Elements in the ring $R_q = \mathbb{Z}_q[X]/(X^n+1)$ is represented as polynomials of degree less than $n$.
In ring arithmetic, modular reduction is a crucial operation that reduces the size of large integers.
The Montgomery reduction \cite{montgomery1985modular} and Barrett reduction \cite{DBLP:conf/crypto/Barrett86} are two techniques employed for fast and constant-time modular reduction, by replacing the time-consuming division with faster multiplication or bit shift operations.

The multiplication of two polynomials $\bm{f} = \sum_{i=0}^{n-1}f_i X^i$, $\bm{g} = \sum_{j=0}^{n-1}g_j X^j$ over the ring $R_q$ results in a polynomial $\bm{h} = \sum_{t=0}^{n-1}h_t X^t \in R_q$, of which the coefficients $(h_0,\ldots,h_{n-1})$ is the negacylic convolution of $(f_0,\ldots,f_{n-1})$ and $(g_0,\ldots,g_{n-1})$.
However, direct multiplication and accumulation of results using the schoolbook method has a computational complexity of $\mathcal{O}(n^2)$, making this operation a major performance bottleneck.
To accelerate polynomial multiplication, the Number-Theoretic Transform (NTT) is commonly used.
NTT is a variant of Discrete Fourier Transform (DFT) that operates on integers modulo a prime number.
By denoting $\psi$ as the primitive $2n$-th root of unity, we can formulate the forward and inverse negacyclic NTT for polynomial $f\in R_q$:
\begin{gather*}
	\hat{\bm{f}}:=\mathsf{NTT}(\bm{f}), \hat{f}_j=\sum_{i=0}^{n-1} f_i \psi^{(2i+1)j} \pmod q \\
	\bm{f}:=\mathsf{INTT}(\hat{\bm{f}}), f_i=\frac{1}{n} \sum_{j=0}^{n-1} \hat{f}_j \psi^{-(2i+1)j} \pmod q
\end{gather*}
Using this technique, we can perform the multiplication as $\bm{f}\cdot\bm{g} := \mathsf{INTT}(\mathsf{NTT}(\bm{f})\cdot\mathsf{NTT}(\bm{g}))$.
This reduces the computational complexity of polynomial multiplication from $\mathcal{O}(n^2)$ to $\mathcal{O}(n\log n)$, as both $\mathsf{NTT}$ and $\mathsf{INTT}$ have a computational complexity of $\mathcal{O}(n\log n)$.

\subsection{\cc-\dd} \label{sec:pre-dili}

\dd\ is a lattice-based digital signature scheme designed with efficiency and practicality, providing security in the Quantum Random Oracle Model (QROM) under the Module Learning With Errors (MLWE) and a variant of the Module Short Integer Solution (MSIS) assumptions \cite{DBLP:journals/tches/DucasKLLSSS18,nist-dili}.
The scheme design is based on the \textit{Fiat-Shamir with Aborts} approach \cite{DBLP:conf/asiacrypt/Lyubashevsky09,DBLP:conf/eurocrypt/Lyubashevsky12}.
Within this framework, a prover aims to demonstrate the knowledge of a secret key without revealing it. The prover and verifier engage in an interactive protocol where the prover generates commitments and responses to random challenges from the verifier. This interaction can be converted to a non-interactive digital signature scheme by generating the challenge through hashing the commitment with the message to be signed.

The scheme comprises three procedures: key generation ($\mathsf{Gen}$), signing ($\mathsf{Sign}$), and verification ($\mathsf{Verify}$), as specified in Algorithm \ref{algo:gen}, \ref{algo:sign}, and \ref{algo:verify}, respectively. 
Table \ref{tab:parameter} presents the parameters of \dd\ corresponding to three NIST security levels. These parameters have been carefully chosen to balance security, efficiency, and practicality, making Dilithium a viable choice for various real-world applications. As a lattice-based digital signature scheme, \dd\ exhibits robust post-quantum security properties that make it suitable for long-term deployment in the rapidly evolving field of cryptography.
Below are some specifications of the components used in this scheme.

\noindent
\begin{minipage}{0.49\textwidth}
\begin{algorithm}[H]
    \centering
    \caption{\dd.$\mathsf{Gen}$}\label{algo:gen}
    \footnotesize
    \begin{algorithmic}[1]
        \Ensure $pk = (\rho, \mathbf{t}_1), sk = (\rho, K, tr, \mathbf{s}_1, \mathbf{s}_2, \mathbf{t}_0)$

        \State $ \zeta \leftarrow\{0,1\}^{L_1} $
        \State $ (\rho, \rho^{\prime}, K) \in\{0,1\}^{L_1\times L_2 \times L_1} := \mathcal{H}(\zeta) $
        \State $ \hat{\mathbf{A}} \in R_q^{k \times \ell} := \mathsf{ExpandA}(\rho) $
        \State $ (\mathbf{s}_1, \mathbf{s}_2) \in S_\eta^\ell \times S_\eta^k := \mathsf{ExpandS}(\rho^{\prime}) $
        \State $ \mathbf{t} := \mathsf{INTT}(\hat{\mathbf{A}} \cdot \mathsf{NTT}(\mathbf{s}_1))+\mathbf{s}_2 $
        \State $ (\mathbf{t}_1, \mathbf{t}_0) := \mathsf{Power2Round}_q(\mathbf{t}, d) $
        \State $ tr \in\{0,1\}^{L_1} := \mathcal{H}(\rho \| \mathbf{t}_1) $
        \State $ pk := (\rho, \mathbf{t}_1)$
        \State $sk := (\rho, K, tr, \mathbf{s}_1, \mathbf{s}_2, \mathbf{t}_0)$
    \end{algorithmic}\end{algorithm}
\vspace{-6mm}
\setcounter{algorithm}{2}
\begin{algorithm}[H]
    \centering
    \caption{\dd.$\mathsf{Verify}$}\label{algo:verify}
    \footnotesize
    \begin{algorithmic}[1]
        \Require $pk = (\rho, \mathbf{t}_1), M \in\{0,1\}^*, \sigma=(\tilde{c},\mathbf{z}, \mathbf{h})$
        \Ensure $ r $
        
        \State $ \hat{\mathbf{A}} \in R_q^{k \times \ell}:=\mathsf{ExpandA}(\rho) $
        \State $ \mu \in\{0,1\}^{L_2}:=\mathcal{H}\left(\mathcal{H}\left(\rho \| \mathbf{t}_1\right) \| M\right) $
        \State $ \bm{c}:=\mathsf{SamplelnBall}(\tilde{c}); \hat{\bm{c}}:=\mathsf{NTT}(\bm{c}) $
        \State $ \mathbf{v} := \mathsf{INTT}(\hat{\mathbf{A}}\cdot\mathsf{NTT}(\mathbf{z})-\hat{\bm{c}}\cdot\mathsf{NTT}(\mathbf{t}_1 \cdot 2^d)) $
        \State $ \mathbf{w}_1^{\prime}:=\mathsf{UseHint}_q\left(\mathbf{h}, \mathbf{v}, 2 \gamma_2\right) $
        \State $ r := \llbracket\|\mathbf{z}\|_{\infty}<\gamma_1-\beta\rrbracket \& \llbracket\tilde{c}=\mathcal{H}(\mu \| \mathbf{w}_1^{\prime}) \rrbracket \& \llbracket[ \mathbf{h} ] \leq \omega\rrbracket $
    \end{algorithmic}
\end{algorithm}
\end{minipage}
\hfill
\begin{minipage}{0.49\textwidth}
\setcounter{algorithm}{1}
\begin{algorithm}[H]
    \centering
    \caption{\dd.$\mathsf{Sign}$}\label{algo:sign}
    \footnotesize
    \begin{algorithmic}[1]
        \Require $ sk=(\rho, K, t r, \mathbf{s}_1, \mathbf{s}_2, \mathbf{t}_0) $, $ M \in\{0,1\}^* $
        \Ensure $ \sigma=(\tilde{c},\mathbf{z}, \mathbf{h}) $
        
        \State $ \mu \in\{0,1\}^{L_2}:=\mathcal{H}(tr \| M)$
        \State $ \rho^{\prime} \in\{0,1\}^{L_2}:=\mathcal{H}(K \| \mu) $
        \State $ \hat{\mathbf{A}} \in R_q^{k \times \ell}:=\mathsf{ExpandA}(\rho); \kappa:=0 ;(\mathbf{z}, \mathbf{h}):=\perp $
        \State $ \hat{\mathbf{s}}_1:=\mathsf{NTT}(\mathbf{s}_{1}) ; \hat{\mathbf{s}}
        _2:=\mathsf{NTT}(\mathbf{s}_{2}) ; \hat{\mathbf{t}}_0:=\mathsf{NTT}(\mathbf{t}_0) $
        \While {$ (\mathbf{z}, \mathbf{h})=\perp $}
        	\State $ \mathbf{y} \in S_{\gamma_1}^{\ell}:=\mathsf{ExpandMask}(\rho^{\prime}, \kappa) $
        	\State $ \mathbf{w}:=\mathsf{INTT}(\hat{\mathbf{A}} \cdot \mathsf{NTT}(\mathbf{y})) $
        	\State $ \mathbf{w}_1:=\mathsf{HighBits}_q(\mathbf{w}, 2\gamma_2) $
        	\State $ \tilde{c} \in \{0,1\}^{L1} := \mathcal{H}(\mu \| \mathbf{w}_1)$
        	\State $ \bm{c} \in B_\tau :=\mathsf{SamplelnBall}(\tilde{c}); \hat{\bm{c}}:=\mathsf{NTT}(\bm{c})$
        	\State $ \mathbf{z}:=\mathbf{y}+\mathsf{INTT}(\hat{\bm{c}} \cdot \hat{\mathbf{s}}_1) $
        	\State $ \mathbf{v}_s:= \mathsf{INTT}(\hat{\bm{c}} \cdot \hat{\mathbf{s}}_2)$
        	\State $ \mathbf{r}_0:=\mathsf{LowBits}_q(\mathbf{w}-\mathbf{v}_s, 2\gamma_2) $
        	\If {$\|\mathbf{z}\|_{\infty} \geq \gamma_1-\beta \textbf { or }\|\mathbf{r}_0\|_{\infty} \geq \gamma_2-\beta$}
        		\State $ (\mathbf{z}, \mathbf{h}):=\perp $
        	\Else
        		\State $ \mathbf{v}_t:= \mathsf{INTT}(\hat{\bm{c}}\cdot \hat{\mathbf{t}}_0)$
        		\State $ \mathbf{h}:=\mathsf{MakeHint}_q(-\mathbf{v}_t, \mathbf{w}-\mathbf{v}_s+ \mathbf{v}_t, 2 \gamma_2) $
        		\If {$ \|\mathbf{\mathbf{v}_t}\|_{\infty} \geq \gamma_2 \text { or }  [ \mathbf{h} ] > \omega$}
        		\State $ (\mathbf{z}, \mathbf{h}):=\perp $
        		\EndIf
        	\EndIf
        	\State $ \kappa:=\kappa+\ell $
        \EndWhile
    \end{algorithmic}
\end{algorithm}
\end{minipage}

\begin{table}[H]
\setlength{\tabcolsep}{3.0mm}
\centering
\caption{Parameter specifications of \dd\ for three NIST security levels}
\label{tab:parameter}
\begin{tabular}{c|cccccccccccc}
\hline
Level & $n$ & $q$     & $(k,\ell)$ & $d$ & $\tau$ & $\gamma_1$ & $\gamma_2$ & $\eta$ & $\beta$ & $\omega$ & $L_1$ & $L_2$ \\\hline
2     & 256 & 8380417 & (4,4)   & 13  & 39     & $2^{17}$   & $95232$ & 2      & 78      & 80 & 256 & 512       \\
3     & 256 & 8380417 & (6,5)   & 13  & 49     & $2^{19}$   & $261888$ & 4      & 196     & 55 & 256 & 512       \\
5     & 256 & 8380417 & (8,7)   & 13  & 60     & $2^{19}$   & $261888$ & 2      & 120     & 75 & 256 & 512   \\\hline   
\end{tabular}
\end{table}

\paragraph{Sampling.} \label{sec:pre-sampling}
In \dd, SHAKE128/256 \cite{sha3} is used as the extendable-output function (XOF) to extend input seeds into a sufficient number of random bytes, and SHAKE256 is employed to instantiate the hash function $\mathcal{H}$. 
A \textit{rejection sampling} mechanism is applied to generate sequences that follow a uniform distribution by selecting $|\mathcal{B}|$ bits per sample and retaining only those that are less than $\mathcal{B}$. 
Based on this mechanism, the $\mathsf{ExpandA}$ function produces a matrix $\hat{\mathbf{A}}$ with coefficients in the range $[0, q)$, while the $\mathsf{ExpandS}$ and $\mathsf{ExpandMask}$ functions generate vectors with coefficients in the ranges $[-\eta, \eta]$ and $[-\gamma_1 + 1, \gamma_1]$, respectively. 
Moreover, the $\mathsf{SamplelnBall}$ function generates a sparse polynomial $c$ with $\tau$ nonzero coefficients by iteratively determining $\tau$ valid positions in the polynomial to hold nonzero integers, achieved by comparing a random byte with the current loop index.

\paragraph{Bits Extraction and Hints.}
To reduce communication bandwidth, \dd\ utilizes optimizations that compress both the public key and signature. However, this compression may occasionally lead to verification failures. To address this issue, a hint is incorporated into the signature to prevent failure and ensure the  robustness. The following functions are used to compute high- and low-order bits and the hint.
The $\mathsf{Power2Round}_q(a, d)$ function divides an integer $a$ into $(a_0, a_1):= (a \bmod^{\pm}2^d, (a-a_0)/2^d)$.
The $\mathsf{LowBits}_q(r, \alpha)$ and $\mathsf{HighBits}_q(r, \alpha)$ functions extract the low- and high-order bits $r_1$ and $r_0$, respectively.
By evaluating $\mathcal{P} := (r-r_0=q-1)$, where $r_0:=(r\bmod ^{+} q) \bmod ^{\pm} \alpha$, the outputs are $(r_1,r_0):=(0, r_0-1)$ if $\mathcal{P}$ is true, and $(r_1,r_0):=((r-r_0)/\alpha, r_0)$ otherwise.
The $\mathsf{MakeHint}_q(z,r,\alpha)$ function computes $r_1:=\mathsf{HighBits}_q(r, \alpha)$ and $z_1:=\mathsf{HighBits}_q(r+z, \alpha)$ and returns $\llbracket r_1\neq v_1 \rrbracket$.
The $\mathsf{UseHint}_q(h,r,\alpha)$ function computes $m:=(q-1)/\alpha$ and extracts $(r_1,r_0)$, outputting $r_1$ if $h=0$, otherwise returning $(r_1+1)\bmod^{+}m$ if $r_0>0$ and $(r_1-1)\bmod^{+}m$ if $r_0\leq 0$.
All the above algorithms can be extended to ring elements by applying them coefficient-wise.

\paragraph{Rejection Loop.}
During the signing process, a rejection stage is necessary to prevent the generated signature $\mathbf{z}$ from revealing information about the secret key. The security requirement is not met if $||\mathbf{z}||_\infty \geq \gamma_1-\beta$. Additionally, if any coefficient of the low-order bits of $\mathbf{A}\mathbf{y}-c \mathbf{s}_2$ is greater than $\gamma_2-\beta$, the security and correctness requirements are not satisfied. The loop in the signing procedure is repeated until these conditions are met. The expected number of repetitions is 4.25, 5.1, and 3.85 for the three security levels, respectively.

\subsection{GPU Basics} \label{sec:pre-gpu}

\begin{figure}[t]
\centering
    \includegraphics[scale=1.0]{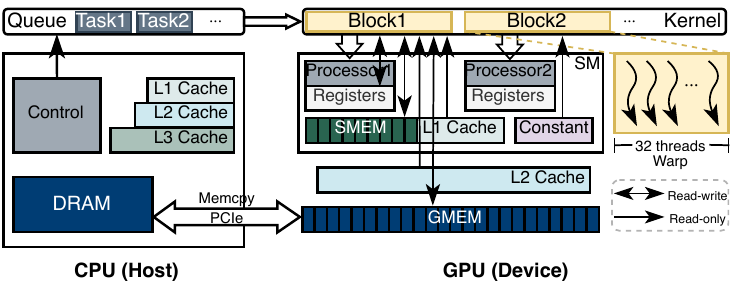}
    \caption{The architecture and host-device computational model.} \label{fig:gpu}
\end{figure}

In the CPU-GPU collaborative computing model, the CPU acts as a host, dispatching kernels to the GPU for execution. Under synchronous computing, the CPU waits for the GPU to complete its tasks, while in asynchronous computing, the CPU continues to perform other tasks. Figure \ref{fig:gpu} illustrates the architecture and computational model.

Modern GPUs facilitate highly parallel kernel processing by employing a massive number of concurrent threads. The CUDA programming model offers a direct interface for accessing hardware resources. Threads are organized into thread blocks, with multiple independent blocks forming a grid. 
GPUs contain various types of memory, including registers, constant memory, shared memory (SMEM), global memory (GMEM), local memory (LMEM), etc. 
GPUs rely on memory hierarchy for low-latency data access. 
Each executing thread has private registers and local memory. Registers provide the fastest access speed, while local memory has the same latency as global memory.
Threads within a block share SMEM, which boasts higher bandwidth and lower latency than either local or global memory. 
GMEM is accessible to all threads but has the highest IO latency and is accessed through the SM L1 cache and GPU L2 cache.

During execution, blocks are assigned to streaming multiprocessors (SM), with every 32 threads grouped into a warp for parallel processing. 
Occupancy, the ratio of active warps per multiprocessor to the maximum possible number of active warps, reflects hardware resource utilization. 
CUDA provides several warp-level primitives. 
In a warp, \textit{warp shuffle} functions allow synchronized register data exchange, \textit{Warp vote} functions enable combining a value in each thread in a tree-reduction pattern and broadcasting the result.

Instructions are executed in warps. 
Optimal instruction throughput occurs when all threads within a warp execute the same instruction; otherwise, threads may diverge which reduces the number of active threads per cycle and performance. 
Instruction of different types are scheduled to different pipelines. 
For example, the Arithmetic Logic Unit (ALU) executes most bit manipulation and logic instructions, and the Load Store Unit (LSU) pipeline mainly issues load and store instructions for memory access.

\section{Implementation Details} \label{sec:imp}

In this section, we illustrate our design rationale and detail our GPU implementation of each operation involved in \dd, encompassing NTT, hash functions, rejection sampling, etc.
By carefully optimizing these operations for the GPU architecture, we aim to leverage the full potential of parallelism and enhance the overall performance. 
Below, we discuss the specific techniques employed for each operation, addressing the challenges encountered in achieving efficient parallelization and memory access, as well as the innovative strategies adopted to overcome these obstacles.

\subsection{Design Rationale}

\dd\ involves extensive linear operations on matrices and vectors with a relatively small modulus, which results in algorithmic simplicity but  high memory overhead. 
Consequently, memory-bound computations are dominated by IO latency. 
Previous works that employ a single thread or partial threads in a warp \cite{DBLP:journals/tpds/GuptaJCC21,DBLP:journals/tpds/GaoXW22} for computation can lead to uncoalesced memory access within a warp, thereby decreasing throughput. 
Another work \cite{seo2022parallel} focuses on warp-level implementation, demonstrating the least overall time when using 32 threads per block for one signing and verification process.

Warp-level implementations offer substantial benefits, including coalesced memory access and significant IO latency reduction. 
Moreover, Dilithium includes numerous sequential computations, such as rejection sampling and number counting, where each thread must be aware of its neighboring threads' states. 
A warp represents the maximum unit within which threads can share their registers.
With these considerations, we first propose a warp-based implementation for \dd. 
In our approach, a key generation, signing, and verification are treated as a single task, with one warp responsible for each task. 
Each block initializes one warp, and each kernel batch processes multiple blocks, thereby allowing several tasks to be processed during each kernel launch. 
This method achieves high throughput, rapid data access speeds, and low amortized time per iteration simultaneously.

However, the resulting SM occupancy under this setting is less than ideal. 
For example, on GPU devices with compute capability 8.6, under an equal partition between L1 cache and shared memory, the theoretical maximum occupancy of each multiprocessor can only reach 33\%, with a maximum warp occupancy of 16\%. 
To achieve theoretical maximum occupancy, each block can allocate up to 3 KB of shared memory and each thread can use a maximum of 128 registers.
Similar constraints apply when using 64 threads per block. 
Conversely, using 3 or 4 warps per block (i.e., 96 or 128 threads per block) can achieve 100\% theoretical occupancy. 
However, 3 warps are not compatible with all \dd\ parameter sets, as 3 is only a factor of the dimension parameter in \dd3. 
Using 4 warps is ideal, providing a maximum of 40 registers per thread and 4 KB of shared memory per block, with decreased occupancy only occurring beyond these limits.

In fact, most arithmetic operations in \dd\ are well-suited for parallelization, and the results from other threads are not required during thread execution. 
In such cases, launching more warps in a block is advantageous, as it can provide up to 3 times higher theoretical maximum occupancy and significantly improve GPU hardware resource utilization.
Based on these insights, we offer an alternative implementation for arithmetic operations using 4 warps per block. 
To simplify our discussion, we refer to this implementation as ``QWarp'', while ``SWarp'' denotes the single-warp-per-block approach.
Combining the two types of implementations offers additional benefits, including flexibility in choosing the most appropriate implementation depending on specific circumstances, thus ensuring optimal performance and compatibility when combined with other operations.


\subsection{Number-Theoretic Transform} \label{sec:imp-ntt}

\begin{figure}[t]
     \centering
     \begin{subfigure}{0.49\textwidth}
         \centering
         \includegraphics[width=\textwidth]{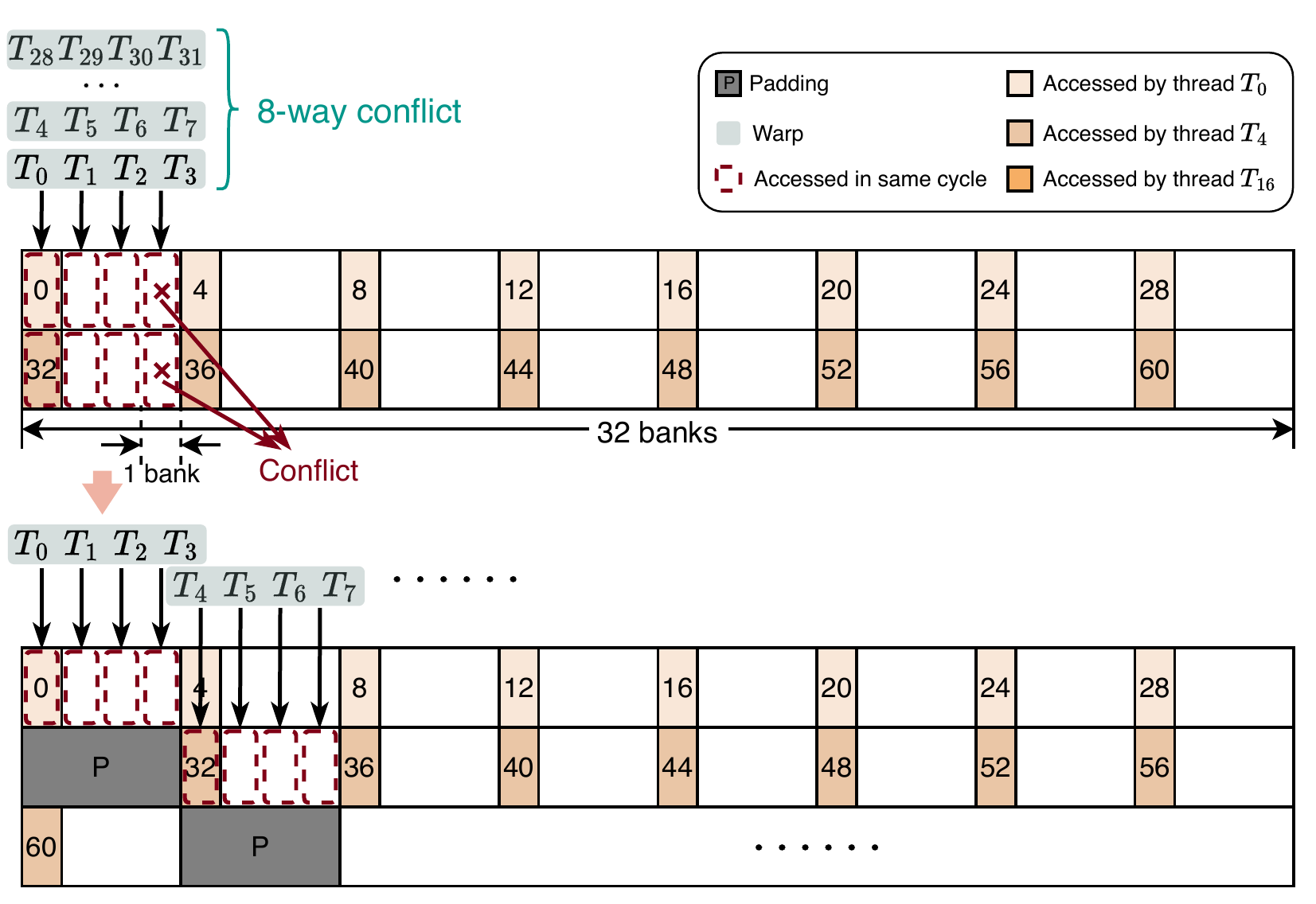}
         \caption{Solving 8-way conflicts between the 3rd and 4th levels in SWarp by padding 4 units every 32 coefficients.}
     \end{subfigure}
     \hfill
     \begin{subfigure}{0.49\textwidth}
         \centering
         \includegraphics[width=\textwidth]{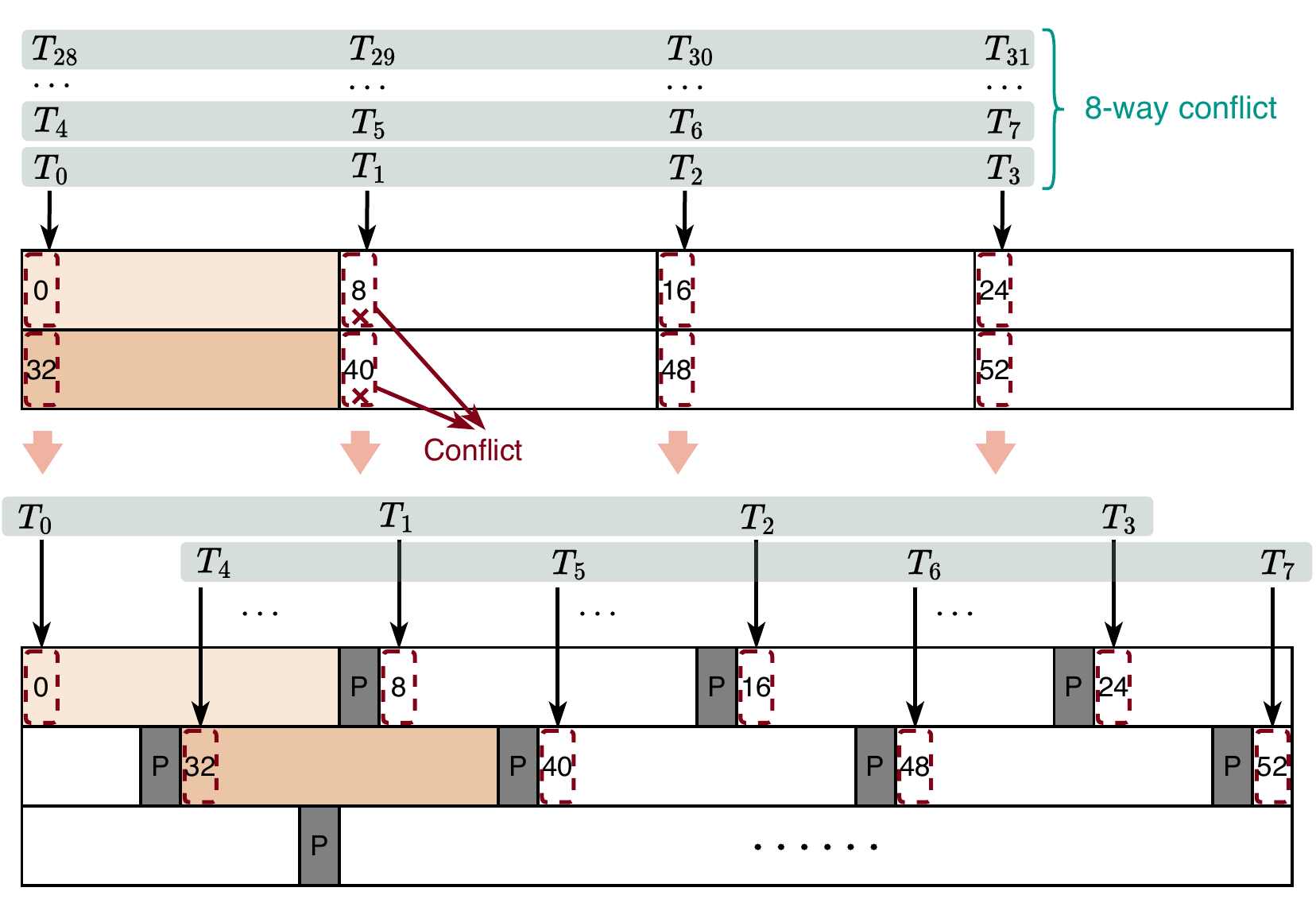}
         \caption{Solving 8-way conflicts between the 6th and 7th levels in SWarp by padding 1 unit every 8 coefficients.}
     \end{subfigure}
     \begin{subfigure}{0.49\textwidth}
         \centering
         \includegraphics[width=\textwidth]{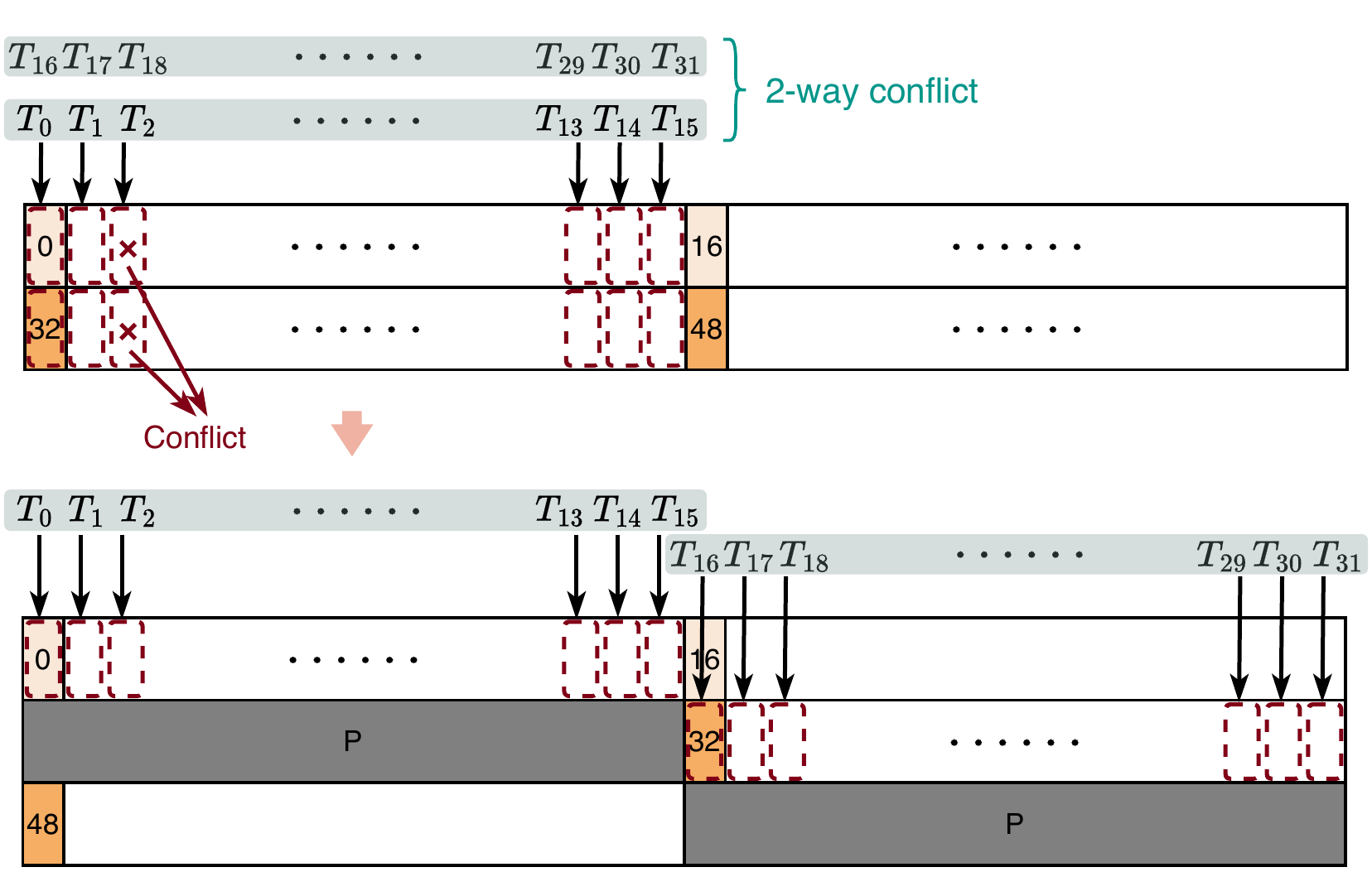}
         \caption{Solving 2-way conflicts between the 3rd and 4th levels in QWarp by padding 16 units every 32 coefficients.}
     \end{subfigure}
     \hfill
     \begin{subfigure}{0.49\textwidth}
         \centering
         \includegraphics[width=\textwidth]{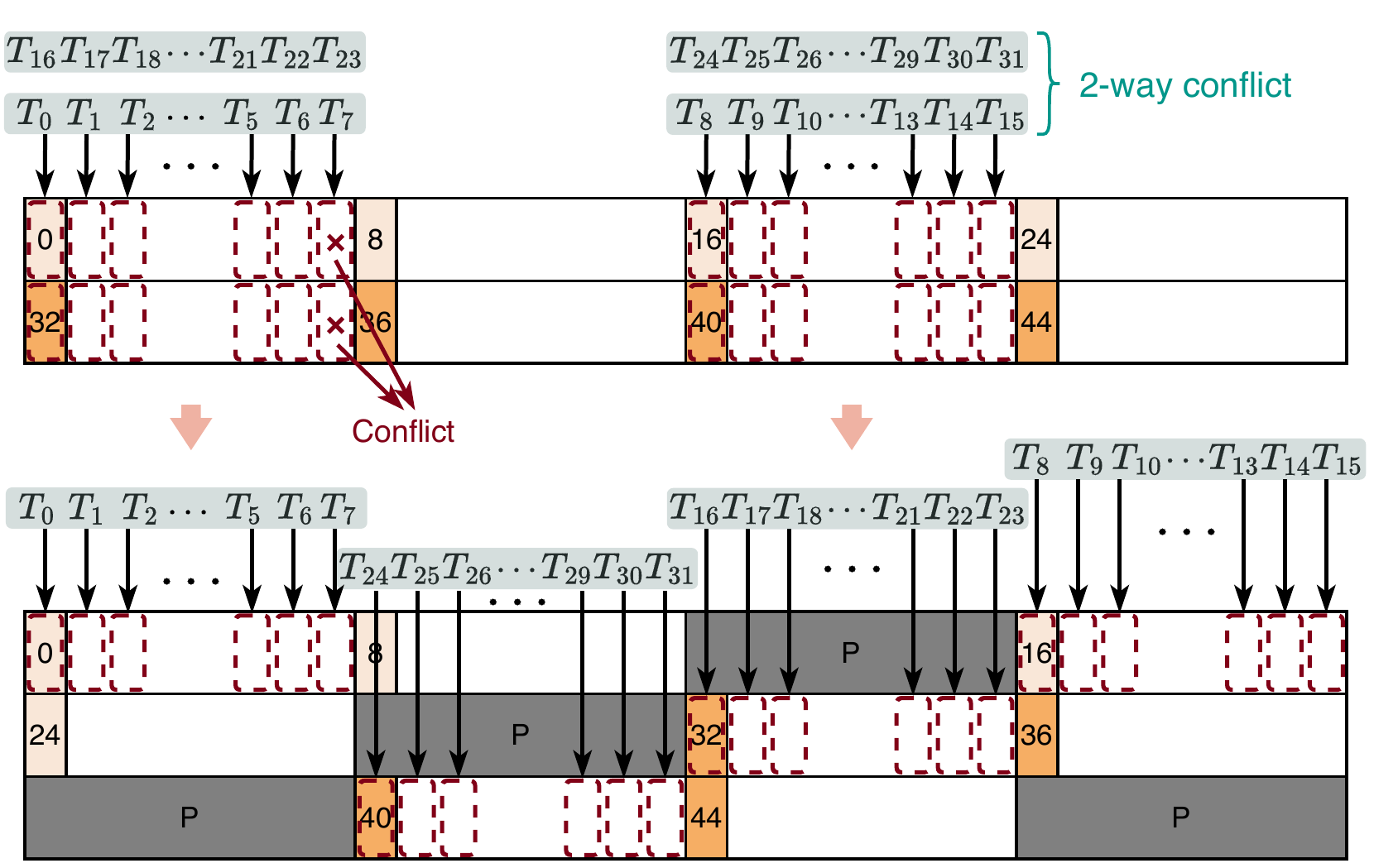}
         \caption{Solving 2-way conflicts between the 4th and 5th levels in QWarp by padding 8 units every 16 coefficients.}
     \end{subfigure}
     \caption{Strategies to solve 8-way and 2-way conflicts in SWarp and QWarp. The SMEM units are organized into 32 memory banks. Bank conflicts occur when multiple threads' requested addresses map to the same memory bank, resulting in serialized accesses.}
     \label{fig:bankconflict}
\end{figure}

We apply the negacyclic NTT with the combination of Cooley-Tukey and Gentleman-Sande algorithms, used in \cite{nist-dili}, to accelerate polynomial multiplication.
Transforming an $n$-dimensional polynomial requires $\log n$ levels.
Since \dd\ fixes the dimension to $n=256$, we implement in-place constant-time 8-level $\mathsf{NTT}$ and $\mathsf{INTT}$.
We exploit both single-warp and multi-warp approaches to match our design for two scenarios.

In the multi-warp approach, we utilize 128 threads for each $\mathsf{NTT}$/$\mathsf{INTT}$ and devise a 2-per-thread implementation that performs a radix-2 butterfly operation on two coefficients at a time.
The distance between the indices of these coefficients is $2^{8-i}$ at level $i$th, where $i\in[1,8]$.
Between levels, we use shared memory to store temporal output from each level for data exchange and prepare input for the next stage.
The single-warp approach utilizes more registers as a trade-off to offer fast data access and reduce IO latency.
Each thread loads eight coefficients into registers and performs a radix-8 $\mathsf{NTT}$/$\mathsf{INTT}$.
This forms merged three-level \cite{DBLP:journals/tches/AlkimBCG20} processing.
We then use SMEM for data exchange and access elements required by following three levels. 
Throughout computation, we only need data transfers between SMEM and registers before the 4th and the 7th levels, where eight loaded coefficients are either continuous or at an interval of 4.
Considering of the high access frequency of the pre-computed table of roots when batch multiple NTT computations, we cache the table in the shared memory to offer low latency load.

To optimize each level individually, we unroll inner loops in both versions while finely tuning execution flow to reduce pipeline stalls.
For multi-warp implementation, since processed coefficients fall within same warp during last five levels of $\mathsf{NTT}$ (resp., first five levels of $\mathsf{INTT}$), we can reduce thread synchronization instructions between these levels.
Additionally, we fuse the multiplications of $n^{-1}$ and roots at last level of $\mathsf{INTT}$ to reduce number of Montgomery multiplications.

Another observation is that strided shared memory accesses for coefficients can lead to bank conflicts where several threads access shared memory units in same bank. 
To prevent stalls from decreasing memory throughput, we pad shared memory carefully so that accesses within a warp fall into individual banks.
Fig. \ref{fig:bankconflict} presents visualization of positions that bank conflicts are triggered in $\mathsf{NTT}$ and our solution to avoid bank conflicts.
Below, we describe only $\mathsf{NTT}$ for simplicity, as the situation is similar for $\mathsf{INTT}$.
In the two data exchange stages of single-warp implementation, eight threads issue instructions to access units in same bank, causing 8-way bank conflicts.
Therefore, we pad four units every 32 coefficients in first exchange stage and one unit every eight coefficients in second exchange stage.
In multi-warp implementation, 2-way conflicts exist at last five levels of $\mathsf{NTT}$ (resp., first five levels of $\mathsf{INTT}$).
At $i$th level ($i\in [3,7]$), we pad $2^{7-i}$ units every $2^{8-i}$ coefficients when writing to shared memory to ensure conflict-free load in next level.
Through these approaches, we reduce overall pipeline stalls while improving compute and memory throughput for both $\mathsf{NTT}$ and $\mathsf{INTT}$.

\subsection{Hash Functions} \label{sec:imp-hash}

We employ warp-level primitives to implement a warp-based SHAKE128/256 following the methodology described in \cite{DBLP:conf/iscas/OnoBS21}.
Although the proposed method is effective, our profiling of their implementation reveals several shortcomings.
It necessitates an excessive number of registers, leading to increased writes to local memory with slow access speeds. 
Furthermore, the global memory access pattern is suboptimal due to the strided and excessive access within the warp. 
Specifically, each thread accesses a 64-bit value through eight consecutive reads (or writes) of 8 bytes each, resulting in a strided global memory access among threads. 
This leads to uncoalesced memory access and increased IO latency. 
Additionally, the excessive memory requests cause high memory input/output (MIO) pipeline utilization, forcing warps to stall and wait for the availability of the MIO instruction queue.

To mitigate these concerns, we optimize the SHAKE implementation by balancing pipeline workloads, optimizing computational flow and memory access pattern, and minimizing resource usage.
In our design, each thread within a warp loads precomputed indices, round constants, and offsets to the registers.
Instead of precomputing and loading all constants into registers \cite{DBLP:conf/iscas/OnoBS21}, we calculate some of the constants during execution and store round constants in constant memory to reduce global memory access.
This approach balances MIO instructions with arithmetic instructions, reducing IO latency and overall stalls, thus improving pipeline performance. 
Second, since each thread needs to load or store 8-byte values during the absorb or squeeze phases, we align input and output streams, replacing the byte-wise load and store operations in \cite{DBLP:conf/iscas/OnoBS21} with a single global memory access intrinsic. 
This optimization results in fewer but wider loads and stores, alleviating pipeline pressure. 
Additionally, we optimize the overall computational flow, particularly the branches responsible for handling insufficient inputs, which reduces thread divergence and enhances overall performance. 
In the permutation function used for internal state updates, the 24 rounds of permutation are executed by 25 threads, and each thread stores 64-bit of the state in registers. In this case, warp-shuffle is employed in each round to enable threads to access the states in other threads. 
The optimizations we have implemented significantly improve overall performance, throughput, and occupancy while substantially reducing memory consumption and access.

\subsection{Rejection Sampling} \label{sec:imp-rej}

\begin{figure}[t]
     \centering
     \begin{subfigure}{0.49\textwidth}
         \centering
         \includegraphics[scale=.7]{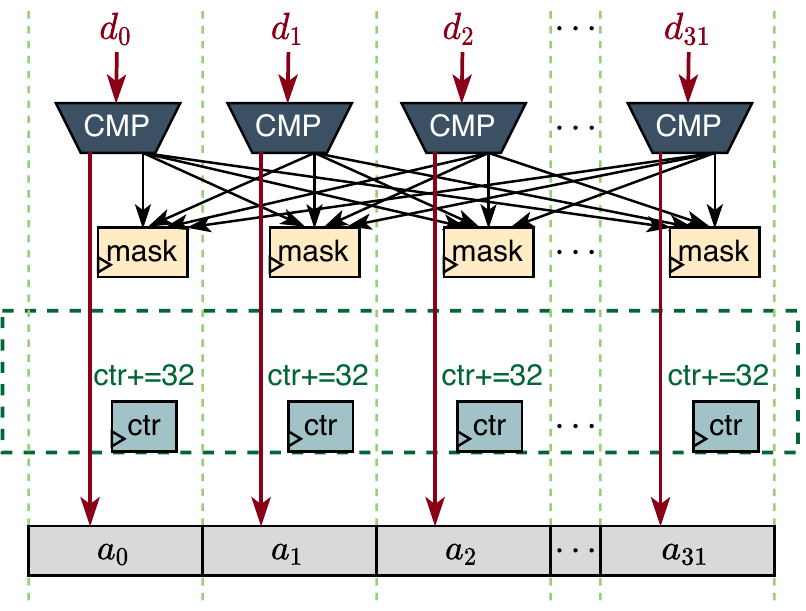}
         \caption{All accept.}
         \label{fig:rejsample-acc}
     \end{subfigure}
     \hfill
     \begin{subfigure}{0.49\textwidth}
         \centering
         \includegraphics[scale=.7]{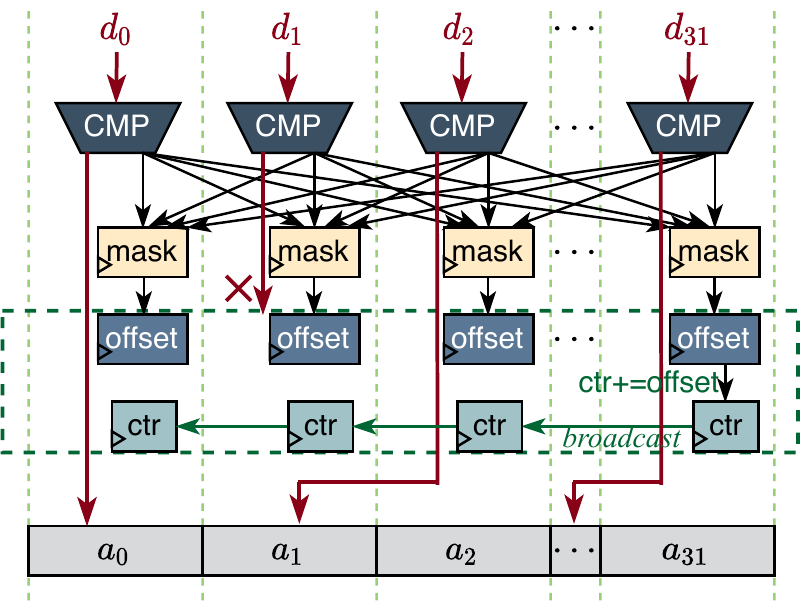}
         \caption{Rejection.}
         \label{fig:rejsample-rej}
     \end{subfigure}
     \caption{Computations in rejection sampling and number counting. Note that number counting does not need to write $d_i$.}
	\label{fig:rejsample}
\end{figure}

\setcounter{algorithm}{3}
\begin{algorithm}[tbp]
    \centering
    \caption{Optimized Rejection sampling in $\mathsf{ExpandA}$}\label{algo:expandA}
    \footnotesize
    \begin{algorithmic}[1]

        \State $ \_\_\mathbf{shared}\_\_ \ \ \mathbf{s}[n], \mathbf{buf}[len] $
        \State $ ctr := 0 $
        \For {$round \in [0,8)$}
        	\State $ pos := round * 32 * 3 + \texttt{threadIdx.x}* 3 $
        	\State $ t := \mathbf{buf}[pos:pos+2] $
        	\State $ sign := [(t-q)\gg 31]\& 1 $ \Comment{Compare with $q$}
        	\State $ mask := \_\_ \mathsf{ballot}\_\mathsf{sync}(\text{0xFFFFFFFF}, sign) $
        	\If {$ mask == \text{0xFFFFFFFF} $} \Comment{All accept}
        		\State $ \mathbf{s}[ctr + \texttt{threadIdx.x}] := t $
        		\State $ ctr := ctr + 32 $
        	\Else {} \Comment{Reject}
        		\State $ mask := mask \ll (31 - \texttt{threadIdx.x}) $
        		\State $ o\textit{ff}set := \_\_\mathsf{popc}(mask) $
        		\If {$(ctr + o\textit{ff}set \leq n) \& sign$}
        			\State $ \mathbf{s}[ctr + o\textit{ff}set-1] = t $
        		\EndIf
        		\State $ ctr := ctr + \_\_\mathsf{shfl}\_\mathsf{sync}(\text{0xFFFFFFFF}, o\textit{ff}set, 31) $
        	\EndIf
        \EndFor
        \State $ \_\_\mathsf{syncwarp}() $
    \end{algorithmic}
\end{algorithm}

Parallelizing massive sequential computations such as rejection sampling and number counting is a major challenge in accelerating \dd.
These computations require comparing elements with a pre-set bound, storing valid ones, and counting the number of valid elements.
The results of these operations depend on the validity of previous ones, which introduces data-dependent conditional computation and unbalanced workload that can lead to thread divergence if not implemented finely.

Prior work \cite{seo2022parallel} precomputes a look-up table to record positions of rejected values and replaces rejection sampling with loading elements directly based on this table during signing and verification. 
While this approach does mitigate the problem, it comes at the cost of increased memory overhead. 
Moreover, it requires additional inputs for the table and transfer it to the verifier along with the public key and signature.
This deviates from the original design intention of reducing communication bandwidth and may cause compatibility issues in applications.

To address this issue, we utilize CUDA integer intrinsics and warp-level primitives in our implementation to ensure that all threads compute the same task. 
Algorithm \ref{algo:expandA} provides our implementation of rejection sampling in $\mathsf{ExpandA}$ function, other functions such as $\mathsf{ExpandS}$ is similar to this.
We set a predicate argument in each thread to record the comparison result so that the entire state of the warp can be obtained through warp voting. 
If all 32 arguments are valid like Fig. \ref{fig:rejsample-acc}, no additional computation is required; otherwise, each thread computes its local inclusive state by an integer intrinsic function to get the offset of the write address as in Fig. \ref{fig:rejsample-rej}.
As only the local counter of the last thread captures correct total numbers, we use warp shuffle synchronization to broadcast it to all threads for correctness in subsequent iterations.

\subsection{Memory-Efficient Inner-Product Computation} \label{sec:imp-innerproduct}

The three procedures in \dd\ require computing the inner-product of a $k\times \ell$ matrix $\mathbf{A}$ with a $\ell \times 1$ vector, denoted as $\mathbf{A} \mathbf{s}_1$, $\mathbf{A}\mathbf{y}$, and $\mathbf{A}\mathbf{z}$ for $\mathsf{Gen}$, $\mathsf{Sign}$, and $\mathsf{Verify}$, respectively. 
Since the matrix is sampled and stored in NTT representation, this operation entails transforming the vector to the NTT domain, performing point-wise multiplication and accumulation, and then transforming back to the normal domain for further computation. 
However, the choice of computational flow and memory type for storing elements affects memory consumption, IO latency, and SM occupancy, which in turn significantly impacts performance.

There are two main computational flows in this process: row-major computation and column-major computation.
Row-major computation involves caching the entire vector of dimension $\ell$ in NTT domain, while column-major computation requires storing an accumulator of dimension $k$.
Since one polynomial in \dd\ comprises of $n$ 32-bit integers, which can be stored as 1 KB in global memory or shared memory, or using 8 registers per thread in SWarp and 2 registers per thread in QWarp.
To optimize performance, we balance between occupancy and IO latency and use different approaches for different scenarios.
Specifically, to compute $\mathbf{A} \mathbf{s}_1$ in $\mathsf{Gen}$ and $\mathbf{A}\mathbf{z}$ in $\mathsf{Verify}$, we apply column-major approach with the on-the-fly computing technique.
In the $\mathsf{Sign}$ function, we use row-major approach for SWarp and column-major approachefor QWarp implementations, respectively.
We describe the details of each implementation below.

\begin{figure}[t]
     \centering
     \begin{subfigure}{1\textwidth}
         \centering
         \includegraphics[scale=.7]{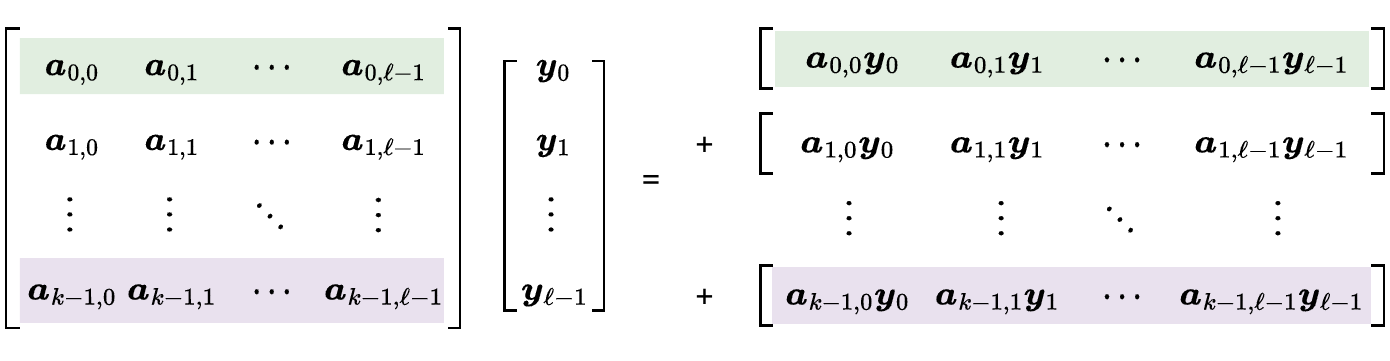}
         \caption{Row-major computation of $\mathbf{Ay}$ in $\mathsf{Sign}$.}
         \label{fig:mult-row}
     \end{subfigure}
     
     \begin{subfigure}{1\textwidth}
         \centering
         \includegraphics[scale=.7]{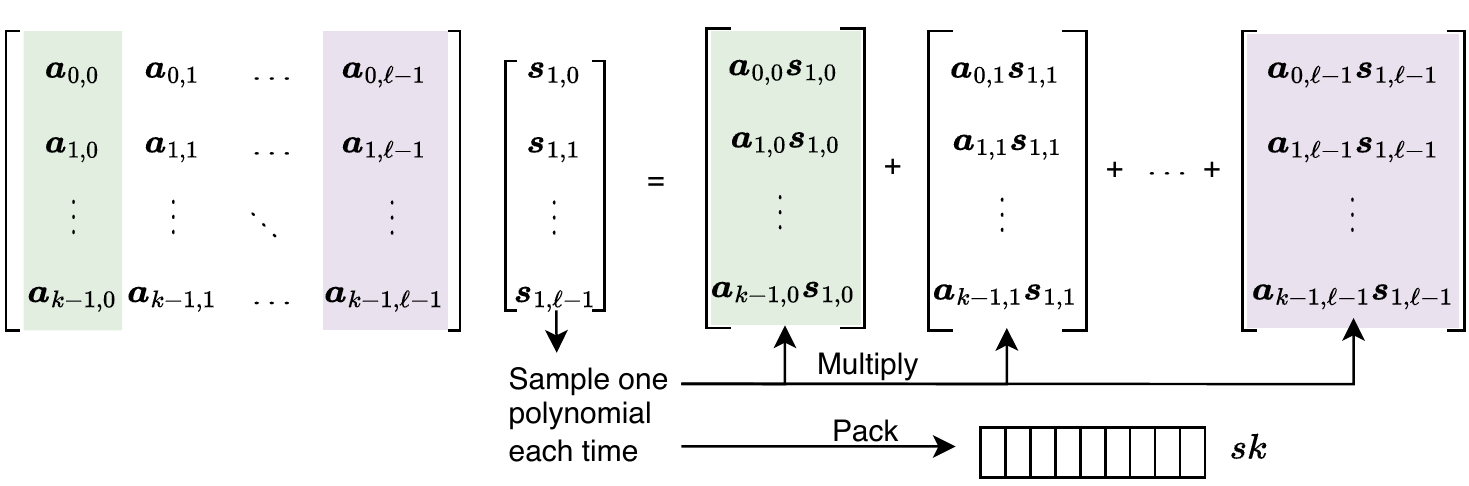}
         \caption{Column-major computation of $\mathbf{A} \mathbf{s}_1$ in $\mathsf{Gen}$.}
         \label{fig:mult-col}
     \end{subfigure}
     \caption{Illustration of inner-product computation in row-major and column-major approaches.}
	\label{fig:mult}
\end{figure}

\paragraph{Computation of $\mathbf{Ay}$ in $\mathsf{Sign}$.}

In SWarp, we apply the row-major approach as shown in Fig. \ref{fig:mult-row}.
We first compute the NTT of $\mathbf{y}$ and store the result in global memory.
Although shared memory provides faster data access, storing $\hat{\mathbf{y}}$ requires $\ell$ KB, which may cause excessive consumption of the shared memory per block and reduce SM occupancy.
Next, we sample each polynomial of $\hat{\mathbf{A}}$, write it into the global memory, and compute the inner-product using 8 registers per thread as accumulators. 
In QWarp, we use a column-major implementation where one polynomial of $\mathbf{y}$ is sampled at a time and its corresponding $\hat{\mathbf{y}}$ is computed using 2 registers per thread for storage. 
We also allocate $2k$ registers as accumulators in each thread for the accumulation results.

\paragraph{On-the-Fly Computing in $\mathsf{Gen}$ and $\mathsf{Verify}$.}

To reduce memory consumption, we adjust the computational flow of sampling of $ \hat{\mathbf{A}} $ and the inner-product computation in both $\mathsf{Gen}$ and $\mathsf{Verify}$. 
Since the multiplication with $ \hat{\mathbf{A}} $ is executed only once in the $\mathsf{Gen}$ and $\mathsf{Verify}$ procedures, it is unnecessary to store the entire matrix, which would consume a significant amount of memory. 
Instead, we generate $ \hat{\mathbf{A}} $ in column order and sample one polynomial of $\bm{s}_{1,j}$ (or unpacked $\bm{z}_j$) at a time, multiplying the corresponding elements and accumulating the results. 
This on-the-fly computation eliminates the need for additional memory allocation beyond a single buffer used as an accumulator to store $k$ polynomials during processing. 
Moreover, by computing polynomials one at a time, we can fuse operations such as sampling, packing, and unpacking into the process, enabling data reuse and reducing the need for data transfers. 
Figure \ref{fig:mult-col} provides an example of row-major computation in the $\mathsf{Gen}$ function and illustrates how we can fuse the packing of the secret key with this process.

\subsection{Memory Pool} \label{sec:imp-mempool}

\begin{figure}
\centering
    \includegraphics[scale=0.8]{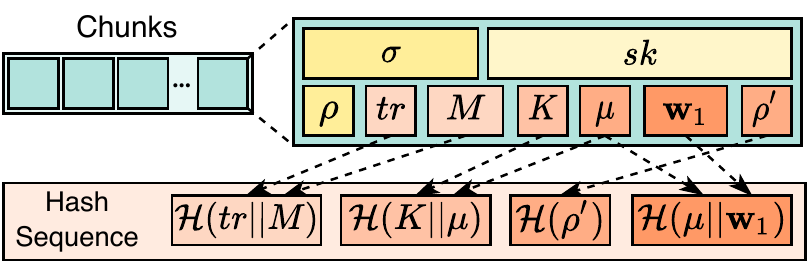}
    \caption{The implemented task-level memory pool (example in $\mathsf{Sign}$) that stores the elements needed in a task processing.} \label{fig:pool}
\end{figure}

In order to ensure efficient memory access during concurrent task processing, we implement a task-level memory pool mechanism with a fixed size. 
The mechanism implemented addresses two important considerations: the storage order of elements and alignment requirements.

First, we carefully arrange the order of elements for both signing and verification processes in accordance with the flow of hash functions. This arrangement guarantees contiguous addresses for input byte streams, as depicted in Fig. \ref{fig:pool}, thereby eliminating the overhead associated with stream concatenation.
Second, we ensure that alignment requirements are met by employing pitch allocation to allocate linear memory for storing seeds and streams. 
Moreover, since the granularity of memory requests to L2, such as an L1TEX request, is a 128-byte cache line (comprising 4 consecutive 32-byte sectors per L2 request), we align the streams to 256 bytes, which include the combinations of seeds for the hash calls. 
This alignment improves the memory request pattern to the L2 cache line.

By implementing this mechanism, we significantly reduce the costs arising from frequent memory allocation and deallocation during computation, resulting in improved performance. This approach further contributes to the optimization of memory management, enabling our implementation to efficiently handle multiple tasks in parallel while maintaining a high level of performance.

\section{Optimizations for Task-level Resource Usage} \label{sec:opt}

In this section, we present a series of optimizations aimed at improving memory and resource usage for our implementation, addressing potential bottlenecks and ensuring efficient utilization of hardware resources. 
We first analyze the profiling results, identifying the bottlenecks. 
Next, we introduce a finely-tuned fusing strategy that takes into account the correlation between operations and adapts the kernels accordingly. 
Following this, we explore dynamic task scheduling to maximize parallelism and further optimize resource usage. 
Finally, we discuss asynchronization and streaming techniques that help to minimize latency and improve overall performance. 
By implementing these optimizations, we are able to achieve better performance and efficient utilization of both memory and computational resources in our implementation.

\subsection{Profiling Results and Analysis}

We profiled the implemented kernels using NVIDIA Nsight Compute and System tools, which provided insights into resource usage and identifies several challenges for improvement. 
Below, we discuss the key observations from our analysis.

\paragraph{High Consumption of SHAKE.}
Despite being well-optimized, the SHAKE implementation still exhibits high resource consumption. 
First, warp-level processing leads to a high number of registers per thread, as each thread is required to load multiple elements. 
However, using other types of memory could increase latency. 
Second, there is a high volume of memory requests, leading to increased pipeline pressure. 
Third, the occupancy is not satisfactory. 
Although it is possible to improve occupancy by batching, for example, four warps in a block where one warp computes a SHAKE, this approach is not easily compatible with the early evaluation technique \cite{DBLP:conf/cardis/RaviGCB19}.

\paragraph{Waste of Hardware Resource in Rejection Loop.}

\begin{figure}[t]
\centering
\includegraphics[scale=0.22]{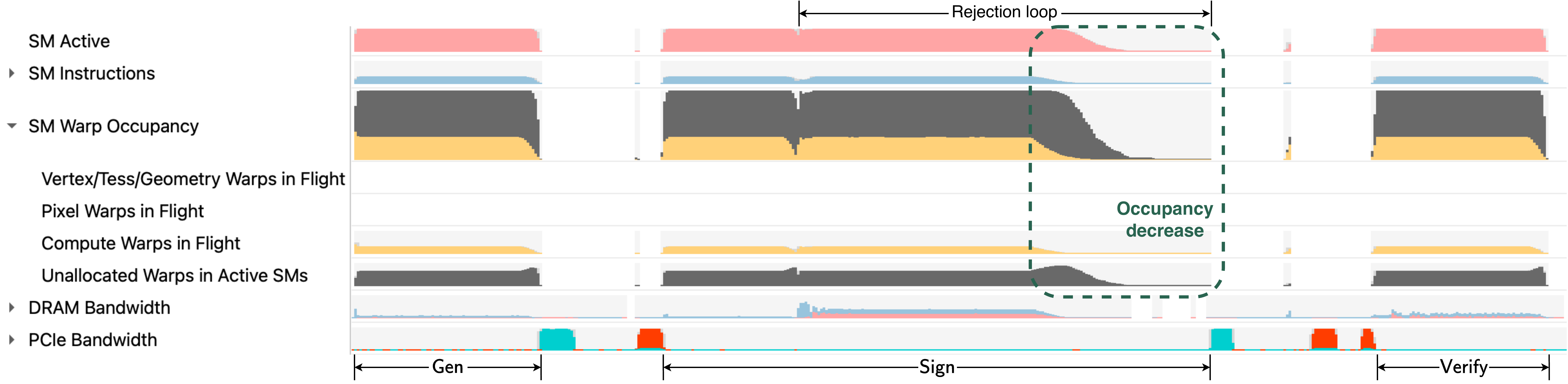}
\caption{Status of the SM during batch execution of multiple tasks (SWarp). During the latter part of the signing process, the hardware resource utilization is insufficient.} \label{fig:sm-occupancy}
\end{figure}

Figure \ref{fig:sm-occupancy} illustrates the streaming multiprocessors (SM) states when batching multiple $\mathsf{Gen}$, $\mathsf{Sign}$, and $\mathsf{Verify}$ tasks. 
The SM occupancy decreases when the processing bar reaches 60\% of the $\mathsf{Sign}$ procedure, revealing that most of the tasks have been completed and the corresponding blocks become idle. 
Although the average number of repeat rounds is around 4, some worse cases may require dozens of rounds, causing many blocks to remain idle while waiting for a very few tasks to finish. 
This waste of hardware resources should be addressed.

\paragraph{Insufficient Use of CPU.}
Synchronization in computation can lead to a waste of CPU computational resources, as it forces tasks to be executed in a predetermined order. 
During the processing, the CPU hosts the kernels to the GPU and waits for the computational results, which gives rise to two issues. 
First, the GPU must wait for the CPU to transfer data, and while waiting for the results, subsequent tasks on CPU cannot proceed until preceding tasks are completed. 
This inefficient use of resources underutilizes the computational capabilities of the CPU, leading to reduced performance. 
Second, synchronization does not allow for hiding the latency of data transfer. The inability to overlap computation and communication is particularly detrimental for GPU computing, where data transfers between the CPU and GPU can become a significant bottleneck. 
This highlights the importance of addressing these inefficiencies in order to fully harness the power of both the CPU and GPU in modern computing systems.

\subsection{Finely-Tuned Fusing Strategy}

While accelerating each operation separately is straightforward, this approach overlooks the correlation between operations and introduces significant kernel launching and data transfer overheads. Since the operations in \dd\ exhibit both internal and external data dependencies, we adapt and fuse the kernels to address these concerns. This technique allows us to store data in registers and shared memory, which offer relatively low IO latency, enabling us to reduce data access by reusing stored data. To avoid over-fusing, which can lead to excessive resource consumption and decreased SM occupancy and performance, we exploit several strategies and carefully fine-tune kernel fusion to optimize performance.

\paragraph{Fusing in SWarp implementation.}
In the SWarp implementation, we fuse the entire $\mathsf{Gen}$, the entire rejection loop in $\mathsf{Sign}$, and the $\mathsf{Verify}$ processes into separate kernels. 
The maximum SM occupancy in the SWarp design can only reach 33\%, so fusing high-consumption kernels, such as SHAKE, with others has little impact on overall occupancy. 
Instead, it provides the advantage of retaining intermediate values in registers and shared memory. 
Combined with early evaluation \cite{DBLP:conf/cardis/RaviGCB19}, this approach significantly reduces overall data read and write requests, particularly time-consuming global memory access. 
In the fused kernels, the majority of memory instructions are concentrated in register and shared memory reading and writing, which have relatively low latency.

\paragraph{Fusing in QWarp implementation.}
In the QWarp implementation, we aim at improving the occupancy of arithmetic operations to accelerate the computation. 
Given the high memory consumption of SHAKE, fusing it with other arithmetic kernels might negatively impact the efficiency of arithmetic operations. 
Therefore, in the $\mathsf{Sign}$ procedure, we implement the rejection loop as four kernels, responsible for sampling $\mathbf{y}$, computing the commitment $\mathbf{w}$ as well as its high- and low-order bits, sampling the challenge polynomial $\bm{c}$, and the final process of computing and checking the validity of signatures, respectively. 
The primary consideration is to separate the computation of SHAKE and arithmetic operations, ensuring that the hash does not significantly reduce occupancy.

\begin{figure}[t]
\centering
    \includegraphics[scale=0.7]{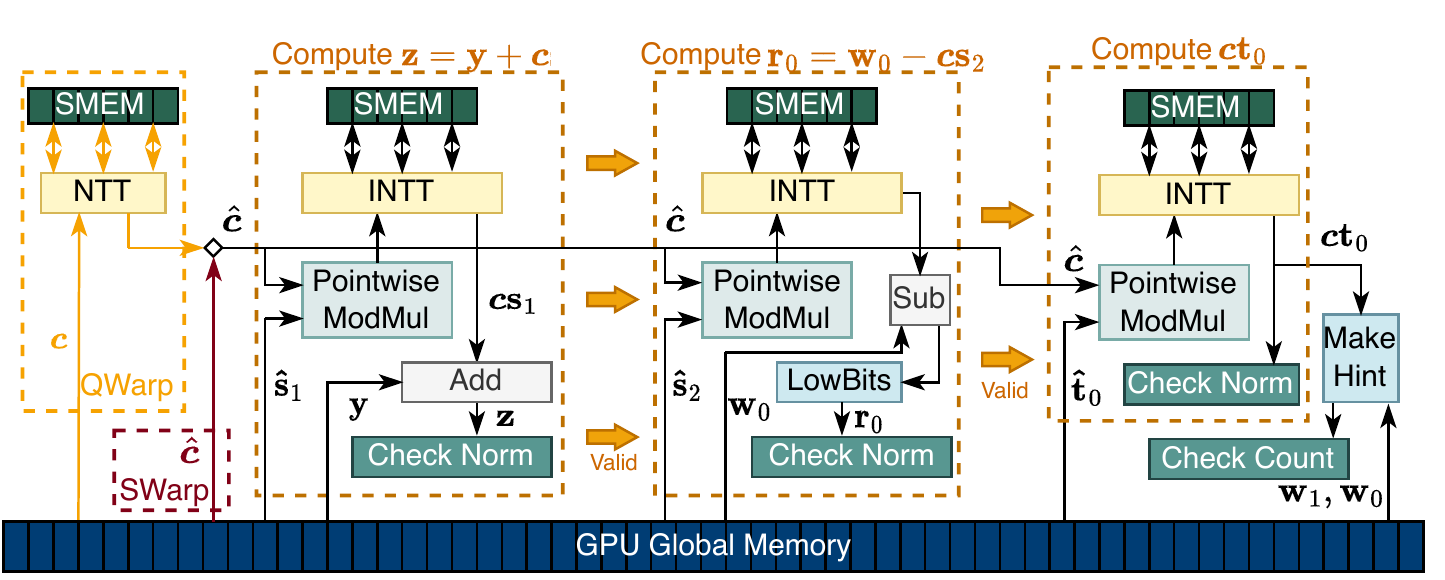}
    \caption{Logical flow and data access pattern in the rejection loop of $\mathsf{Sign}$.} \label{fig:check}
\end{figure}

\paragraph{Fusing in Rejection Loop.}
The computation of the rejection loop in SWarp and QWarp is illustrated in Fig. \ref{fig:check}. 
The two implementations mainly differ in that QWarp uses a fused kernel, whereas SWarp relies on an inline process. 
Additionally, in QWarp, we compute $\hat{c}$ and store it in registers, and in SWarp, $\hat{c}$ is stored in global memory to improve occupancy. 
During processing, the next computation stage is entered only when the $\ell_\infty$-norm of the currently evaluated element is within the preset threshold. 
We apply the early evaluation technique \cite{DBLP:conf/cardis/RaviGCB19} and ensure compatibility between the $\ell_\infty$-norm checking function and arithmetic operations. 
Consequently, in each iteration, we perform computation polynomial-wise and immediately check the coefficients as they are generated. 
This technique enables a more timely rejection.

\subsection{Dynamic Task Scheduling}

\begin{figure}[t]
\centering
    \includegraphics[scale=0.38]{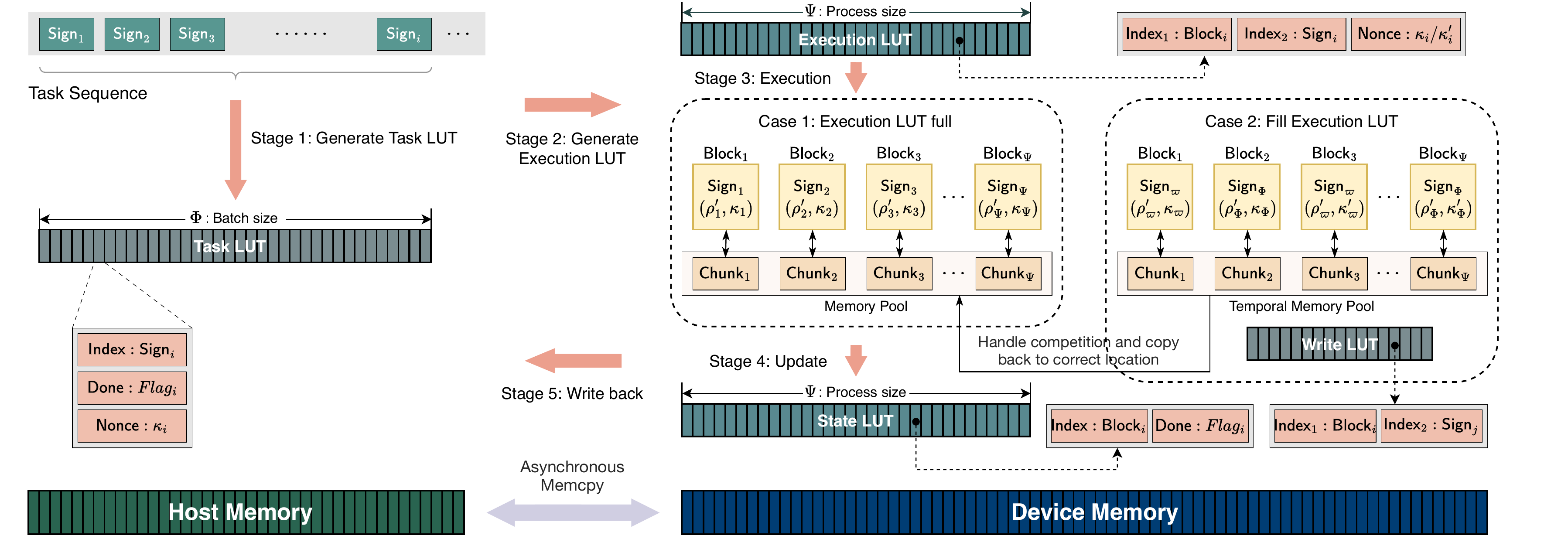}
    \caption{The structure of the dynamic task scheduler. The left side represents the CPU (host), and the right side represents the GPU (device).} \label{fig:scheduler}
\end{figure}

The mathematical expectation of repetitions in the $\mathsf{Sign}$ of \dd\ is around 4, however, in some worse cases, the number of repetitions may reach dozens of times.
When batching multiple $\mathsf{Sign}$ tasks, some blocks become idle after returning valid signatures, however, the computation does not finish until the task with the largest number of repetition rounds is complete. 
This results in wasted GPU hardware resources and low utilization.

To address this issue, we propose a dynamic scheduling strategy that is general and can be applied to digital signature schemes based on the Fiat-Shamir with Aborts approach. 
The main idea of this strategy is to predicate the nonce $\kappa$, and use the execution units which may become idle in the next round to compute this prediction.
This allows us to obtain results of different repetitions under the same seed but with different nonces in one round.
The final output is determined by selecting the valid signature with the smallest nonce.

The design causes a mismatch between the original signature task sequence and its execution sequence on the GPU.
To ensure efficient transfer of all computational results to the CPU, we must maintain the same stored order in the memory pool as in the original task sequence. 
This will allow for a single device-to-host data transfer instruction with relatively low latency.
To achieve this, we construct several maps during scheduling process along with corresponding look-up tables (LUT). 
In our setting, sign tasks are kept in a queue and every time we batch $\Phi$ tasks and delegate them for computation using $\Psi$ concurrent processing units on the device.

The first map bridges signing tasks to GPU execution units through a task LUT of size $\Phi$ and an execution LUT of size $\Psi$. 
The second map reflects validity of outputs from execution units using a state LUT of size $\Psi$, which tracks states and records nonces. 
The device updates both LUTs each round, and usually no units are idle when execution LUT is full.
When remaining tasks cannot fill up execution LUT completely, we predicate the nonce
Another problem needed to solve is the competition of the execution units, where more than one units output valid signatures but with different nonces.
In this case, we use another map to links the signatures with smaller nonces to the corresponding position in the memory pool, for writing from the temporal memory pool to the original memory pool.

Figure \ref{fig:scheduler} illustrates the computational flow and data structure.
During processing, the data transfers between the host and device are asynchronous.
Our prediction and scheduling strategy significantly reduces the number of execution rounds required. 
Specifically, one block computes one signing task in our implementation.
Therefore, we recommend setting $\Psi$, which represents the number of concurrent processing units on the device, to match the maximum active blocks determined through kernel profiling.

\subsection{Asynchronization and Streaming}

Asynchronization plays a crucial role in improving the overall performance of both CPU and GPU by allowing tasks to be executed concurrently without waiting for the completion of previous tasks. 
This approach effectively hides the latency of memory access and computation, resulting in better throughput and reduced idle time. 
By employing asynchronization techniques and managing data transfers more efficiently, we can maximize resource utilization, increase performance, and unlock the full potential of both CPU and GPU.

In order to fully utilize the computational resources of the GPU, we launch multiple CUDA streams, where each stream executes a partition of the entire task set using the dynamic task scheduling technique described above. 
To enable CUDA streams, we asynchronously manage data transfers between the CPU and GPU. 
However, dynamic task scheduling requires synchronization between the CPU and GPU, as the CPU acts as a scheduler to allocate tasks.
To alleviate the impact of synchronization and effectively hide the latency of data transfers, we employ multiple threads on the CPU side. 
Each thread is assigned a unique CUDA stream and executes a partition of the entire task set. 
This approach not only ensures that the CPU and GPU can work concurrently and efficiently but also greatly diminishes the latency of data transfers, leading to improved performance and resource utilization.

\section{Experimental Results} \label{sec:eva}

\subsection{Experimental Setup} \label{sec:eva-setup}

We compile the C/C++ code using g++ 12.2.0 and the GPU implementations with CUDA 11.8 on an Arch Linux system with kernel 5.15. Our implementation is deployed and tested on three GPUs: a NVIDIA Tesla A100 80G PCIe, a NVIDIA GeForce RTX 3090 Ti, and a NVIDIA Tesla V100S PCIe.
Note that our A100 GPU is put in a X8 PCIe slot, while the other two GPUs are put in X16 PCIe slots.
These GPUs represent a wide range of hardware specifications and capabilities, enabling a thorough evaluation of our approach. The performance of the CPU baseline is obtained on an Intel(R) Core(TM) i9-12900KS CPU with 16 cores.
For the evaluation of the three procedures, we measure performance in operations per second (OP/s).
The reported performance results are the medians of 100 executions, where each execution batches multiple tasks.
The latency of data transfer between the CPU and GPU is also included in the evaluation.

\subsection{Performance of Individual Operations} \label{sec:eva-performance}

Below, we present the performance of individual operations in our \dd\ implementation.
First, we compare the SHAKE implementation with \cite{DBLP:conf/iscas/OnoBS21}, an open-source, state-of-the-art GPU implementation of KYBER \cite{DBLP:conf/eurosp/BosDKLLSSSS18}.
Next, we demonstrate the impact of our proposed optimization techniques in NTT.
Afterwards, we provide the profiling results for all operations, containing both SWarp and QWarp implementations.

\paragraph{SHAKE.}
Table \ref{tab:shake256-result} presents the profiling results of computing $ \mu :=\mathcal{H}(tr | M)$ using the implementation from \cite{DBLP:conf/iscas/OnoBS21} and our optimized SHAKE256 implementation.
Both the compute and memory throughput of our implementation show a 5.8\% increase compared to \cite{DBLP:conf/iscas/OnoBS21}, resulting in a 14.1\% decrease in execution time.
Moreover, our implementation significantly reduces memory consumption, requiring 37.2\% fewer registers, leading to a 125\% increase in theoretical occupancy and a 118.0\% increase in achieved occupancy.
Our implementation also demonstrates substantial reductions in the number of instructions and L1 cache requests.
The number of global instructions is reduced by 34.2\%, and the local memory usage is completely eliminated.
L1 cache requests for both loads and stores decrease by 60.6\% and 97.7\%, respectively.
Similarly, L2 cache interactions with global memory and L1 cache are reduced by 87.35\% and 94.81\%, respectively.
These optimizations significantly contribute to the overall improved performance and resource utilization of our SHAKE256 implementation.

\begin{table}[]
\centering
\setlength{\tabcolsep}{1.0mm}
\caption{Profiling results of our SHAKE256 implementation on a 3090 Ti GPU, compared with \cite{DBLP:conf/iscas/OnoBS21}. Pipe utilization refers to the utilization of peak instructions executed.}
\label{tab:shake256-result}
\begin{tabular}{crrrrrrr}
\hline
\multicolumn{8}{c}{\textbf{Improved Performance}}                                                                                                                                                                                                                                 \\ \hline
\multicolumn{1}{c|}{}        & \multicolumn{2}{c|}{Throughput (\%)}                              & \multicolumn{1}{c|}{Execution}    & \multicolumn{2}{c|}{Occupancy (\%)}                                 & \multicolumn{2}{c}{Pipe Utilization (\%)}                    \\ \cline{2-3} \cline{5-8} 
\multicolumn{1}{c|}{}        & \multicolumn{1}{c|}{Compute}    & \multicolumn{1}{c|}{Memory}     & \multicolumn{1}{c|}{Time ($\mu$s)}   & \multicolumn{1}{c|}{Theoretical} & \multicolumn{1}{c|}{Achieved}    & \multicolumn{1}{c|}{LSU}        & \multicolumn{1}{c}{ALU}    \\ \hline
\multicolumn{1}{c|}{\cite{DBLP:conf/iscas/OnoBS21}} & \multicolumn{1}{r|}{90.03}      & \multicolumn{1}{r|}{90.03}      & \multicolumn{1}{r|}{121.98}      & \multicolumn{1}{r|}{33.33}       & \multicolumn{1}{r|}{30.43}       & \multicolumn{1}{r|}{91.58}      & 28.68                      \\ \hline
\multicolumn{1}{c|}{Ours}    & \multicolumn{1}{r|}{95.25}      & \multicolumn{1}{r|}{95.25}      & \multicolumn{1}{r|}{104.80}      & \multicolumn{1}{r|}{75}          & \multicolumn{1}{r|}{66.34}       & \multicolumn{1}{r|}{97.65}      & 31.70                      \\
\multicolumn{1}{c|}{}        & \multicolumn{1}{r|}{(\textbf{+5.8\%})}   & \multicolumn{1}{r|}{(\textbf{+5.8\%})}   & \multicolumn{1}{r|}{(\textbf{$-$14.1\%})}   & \multicolumn{1}{r|}{(\textbf{+125\%})}    & \multicolumn{1}{r|}{(\textbf{+118.01\%})} & \multicolumn{1}{r|}{(\textbf{+6.63\%})}  & (\textbf{+10.55\%})                 \\ \hline\hline
\multicolumn{8}{c}{\textbf{Reduced Resource Usage}}                                                                                                                                                                                                                                 \\ \hline
\multicolumn{1}{c|}{}        & \multicolumn{1}{c|}{Register}   & \multicolumn{2}{c|}{Memory Instructions}                           & \multicolumn{2}{c|}{L1/TEX Cache Requests}                          & \multicolumn{2}{c}{L2 Cache (KB)}                            \\ \cline{3-8}
\multicolumn{1}{c|}{}        & \multicolumn{1}{c|}{Numbers}    & \multicolumn{1}{c|}{Global}     & \multicolumn{1}{c|}{Local}       & \multicolumn{1}{c|}{Loads}       & \multicolumn{1}{c|}{Stores}      & \multicolumn{1}{c|}{L1/TEX}     & \multicolumn{1}{c}{Global} \\ \hline
\multicolumn{1}{c|}{\cite{DBLP:conf/iscas/OnoBS21}} & \multicolumn{1}{r|}{78}         & \multicolumn{1}{r|}{410 K}      & \multicolumn{1}{r|}{650 K}       & \multicolumn{1}{r|}{660 K}       & \multicolumn{1}{r|}{430 K}       & \multicolumn{1}{r|}{49645.82}   & 24853.76                   \\ \hline
\multicolumn{1}{c|}{Ours}    & \multicolumn{1}{r|}{49}         & \multicolumn{1}{r|}{270 K}      & \multicolumn{1}{r|}{0}           & \multicolumn{1}{r|}{260 K}       & \multicolumn{1}{r|}{10 K}        & \multicolumn{1}{r|}{2578.82}    & 3142.63                    \\
\multicolumn{1}{c|}{}        & \multicolumn{1}{r|}{(\textbf{$-$37.18\%})} & \multicolumn{1}{r|}{(\textbf{$-$34.15\%})} & \multicolumn{1}{r|}{(\textbf{$-$100.00\%})} & \multicolumn{1}{r|}{(\textbf{$-$60.61\%})}  & \multicolumn{1}{r|}{(\textbf{$-$97.67\%})}  & \multicolumn{1}{r|}{(\textbf{$-$94.81\%})} & (\textbf{$-$87.35\%})                 \\ \hline
\end{tabular}%
\end{table}

\paragraph{Speedup Breakdown.}

\begin{figure}
\centering
    \includegraphics[width=\textwidth]{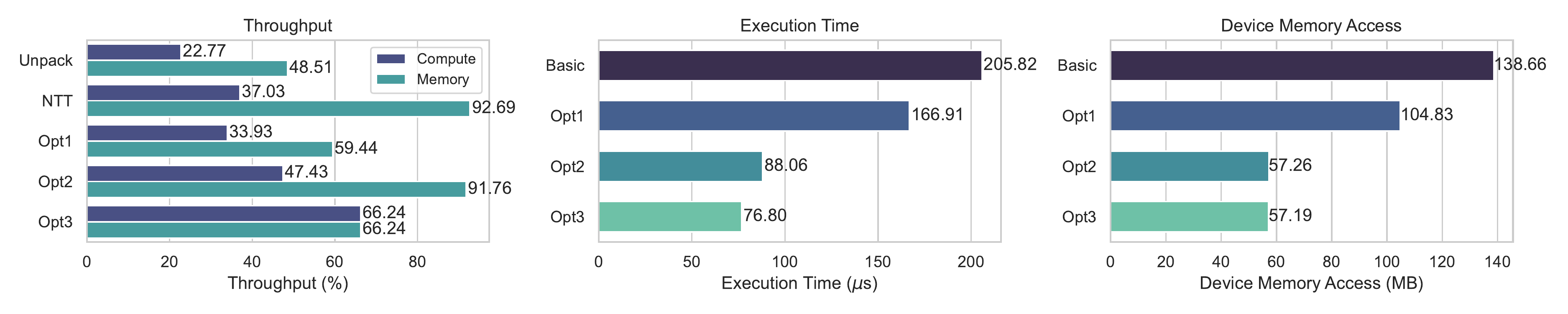}
    \caption{Comparisons of throughput, execution time, and device memory usage between the basic implementation and the step-by-step application of the three optimizations.} \label{fig:ntt-res}
\end{figure}

We use the computation of $\hat{\mathbf{t}}_0$ in the signing procedure as a representative example to demonstrate the impact of the proposed optimizations in our work. This process involves unpacking the secret key to obtain $\mathbf{t}_0$ and then computing $\mathsf{NTT}(\mathbf{t}_0)$. We present the original implementation alongside our step-by-step optimizations, and Fig. \ref{fig:ntt-res} illustrates the throughput results, execution time, and total global memory access for various optimization techniques.
In the basic implementation, we launch an unpacking kernel, store $\mathbf{t}_0$ in GMEM, and subsequently launch another kernel for NTT. We then apply three successive optimizations. The first optimization entails kernel fusion without altering the memory access pattern, leading to an 18.9\% reduction in execution time. The second optimization involves merging loops in the unpacking and NTT processes, using registers to store intermediate values. This results in a 1.4$\times$ improvement in compute throughput, a 47.2\% decrease in execution time, and a 45.4\% reduction in global memory access. Finally, the third optimization addresses bank conflict resolution, producing a kernel with balanced compute and memory throughput and enhancing execution time by 2.7$\times$ compared to the basic implementation.

\paragraph{Comparisons of SWarp and QWarp.}

\begin{table}[t]
\centering
\setlength{\tabcolsep}{3.0mm}
\caption{Performance and profiling results for operations in \dd2 on a 3090 Ti GPU. The ``Type'' column displays the implementation approach and the number of warps in a block, where ``S'' and ``Q'' represent SWarp and Qwarp, respectively. ``The.'' and ``Ach.'' denote theoretical and achieved occupancy, respectively. Memory usage is indicated by the number of registers per thread (``Reg.'') and shared memory (SMEM) in bytes.}
\label{tab:operation-result}
\begin{tabular}{c|r|r|rr|rr|rr}
\hline
Operation                       & \multicolumn{1}{c|}{Type} & \multicolumn{1}{c|}{Time}     & \multicolumn{2}{c|}{Occupancy (\%)}                    & \multicolumn{2}{c|}{Throughput (\%)}                   & \multicolumn{2}{c}{Memory Usage}                     \\ \cline{4-9} 
                                & \multicolumn{1}{c|}{}     & \multicolumn{1}{c|}{($\mu$s)} & \multicolumn{1}{c|}{The.}  & \multicolumn{1}{c|}{Ach.} & \multicolumn{1}{c|}{Com.}  & \multicolumn{1}{c|}{Mem.} & \multicolumn{1}{c|}{Reg.} & \multicolumn{1}{c}{SMEM} \\ \hline
$\mathsf{ExpandA}$                         & (S, 4)                    & 4805.54                       & \multicolumn{1}{r|}{75}    & 67.13                     & \multicolumn{1}{r|}{98.33} & 98.33                     & \multicolumn{1}{r|}{52}   & 3488                     \\ \hline
$\mathsf{ExpandMask}$                      & (S, 1)                    & 1205.25                       & \multicolumn{1}{r|}{33.33} & 32.08                     & \multicolumn{1}{r|}{98.17} & 98.17                     & \multicolumn{1}{r|}{52}   & 752                      \\ \hline
$\mathsf{SampleInBall}$                    & (S, 1)                    & 532.45                        & \multicolumn{1}{r|}{33.33} & 32.00                     & \multicolumn{1}{r|}{97.33} & 97.33                     & \multicolumn{1}{r|}{54}   & 1288                     \\ \hline
\multirow{2}{*}{$\mathsf{NTT}$}            & (S, 1)                    & 22.53                         & \multicolumn{1}{r|}{33.33} & 29.93                     & \multicolumn{1}{r|}{37.50}  & 89.80                      & \multicolumn{1}{r|}{38}   & 1536                     \\ 
                                & (Q, 1)                    & 20.48                         & \multicolumn{1}{r|}{100}   & 90.38                     & \multicolumn{1}{r|}{91.22} & 91.22                     & \multicolumn{1}{r|}{20}   & 2560                     \\ \hline
\multirow{2}{*}{$\mathsf{INTT}$}           & (S, 1)                    & 15.36                         & \multicolumn{1}{r|}{33.33} & 28.57                     & \multicolumn{1}{r|}{58.21} & 61.04                     & \multicolumn{1}{r|}{33}   & 1536                     \\ 
                                & (Q, 1)                    & 20.48                         & \multicolumn{1}{r|}{100}   & 90.62                     & \multicolumn{1}{r|}{90.71} & 90.71                     & \multicolumn{1}{r|}{20}   & 2560                     \\ \hline
\multirow{2}{*}{Inner-product}  & (S, 1)                    & 856.99                        & \multicolumn{1}{r|}{33.33} & 31.77                     & \multicolumn{1}{r|}{27.98} & 76.17                     & \multicolumn{1}{r|}{80}   & 1152                     \\ 
                                & (Q, 1)                    & 346.18                        & \multicolumn{1}{r|}{75}    & 74.10                     & \multicolumn{1}{r|}{68.58} & 80.37                     & \multicolumn{1}{r|}{56}   & 2560                     \\ \hline
\multirow{2}{*}{Rejection loop} & (S, 1)                    & 470.21                        & \multicolumn{1}{r|}{33.33} & 30.41                     & \multicolumn{1}{r|}{41.93} & 74.48                     & \multicolumn{1}{r|}{54}   & 3540                     \\ 
                                & (Q, 1)                    & 299.04                        & \multicolumn{1}{r|}{83.33} & 77.58                     & \multicolumn{1}{r|}{81.93} & 81.93                     & \multicolumn{1}{r|}{48}   & 2688                     \\ \hline
\end{tabular}
\end{table}

Table \ref{tab:operation-result} provides a comprehensive overview of the benchmark and kernel profiling results for our \dd2 implementation, with a detailed comparison of the SWarp and QWarp approaches for various arithmetic operations. In the $\mathsf{ExpandA}$ operation, the optimized hash implementation enables the use of more warps within a block, leading to improved occupancy. This improvement is achieved by sampling four polynomials simultaneously with four warps, effectively increasing the utilization of available resources.
When examining the arithmetic operations, QWarp implementations consistently outperform SWarp across several metrics. For instance, the QWarp approach yields higher occupancy levels, providing a more efficient utilization of GPU resources. Additionally, QWarp delivers more balanced throughput, ensuring a more uniform performance across different operations. It is important to note that the increased shared memory usage in QWarp implementations is a trade-off for avoiding bank conflicts through additional padding. Despite this increase in memory usage, the overall occupancy remains unaffected, as it stays within the acceptable limitations imposed by the hardware.
In conclusion, our \dd2 implementation demonstrates a comprehensive optimization, closely approaching the theoretical maximum occupancy. This highlights the success of our proposed optimizations in addressing the limitations of the original implementation, ultimately leading to enhanced performance, throughput, and occupancy.

\subsection{Sensitivity Study}

\begin{figure}[t]
\centering
    \includegraphics[scale=0.4]{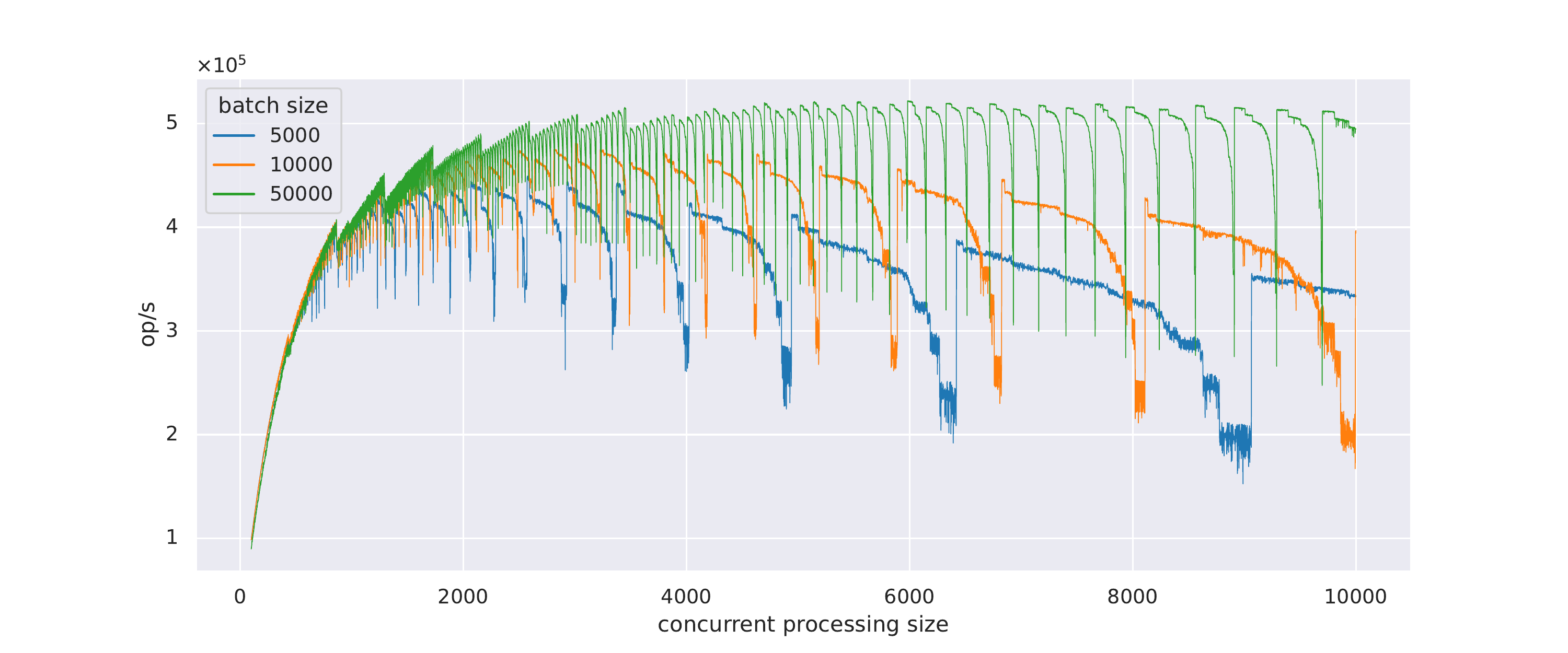}
    \caption{Sensitivity to the concurrent processing sizes ($\Psi$) under three batch sizes ($\Phi$).} \label{fig:res-threshold}
\end{figure}

\begin{figure}[t]
     \centering
     \begin{subfigure}{0.48\textwidth}
         \centering
         \includegraphics[width=0.8\textwidth]{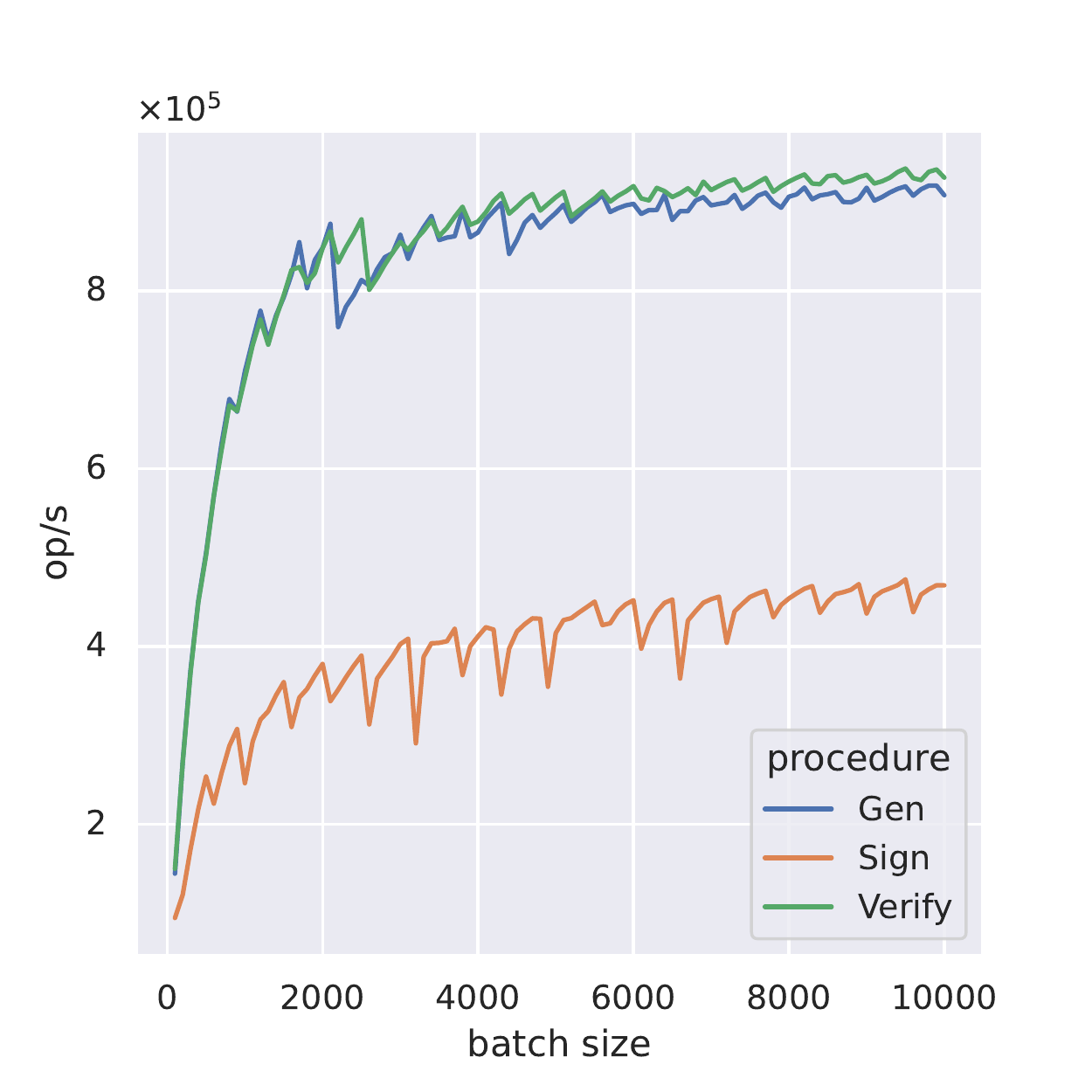}
         \caption{}
         \label{fig:res-batch}
     \end{subfigure}
     \hfill
     \begin{subfigure}{0.48\textwidth}
         \centering
         \includegraphics[width=0.8\textwidth]{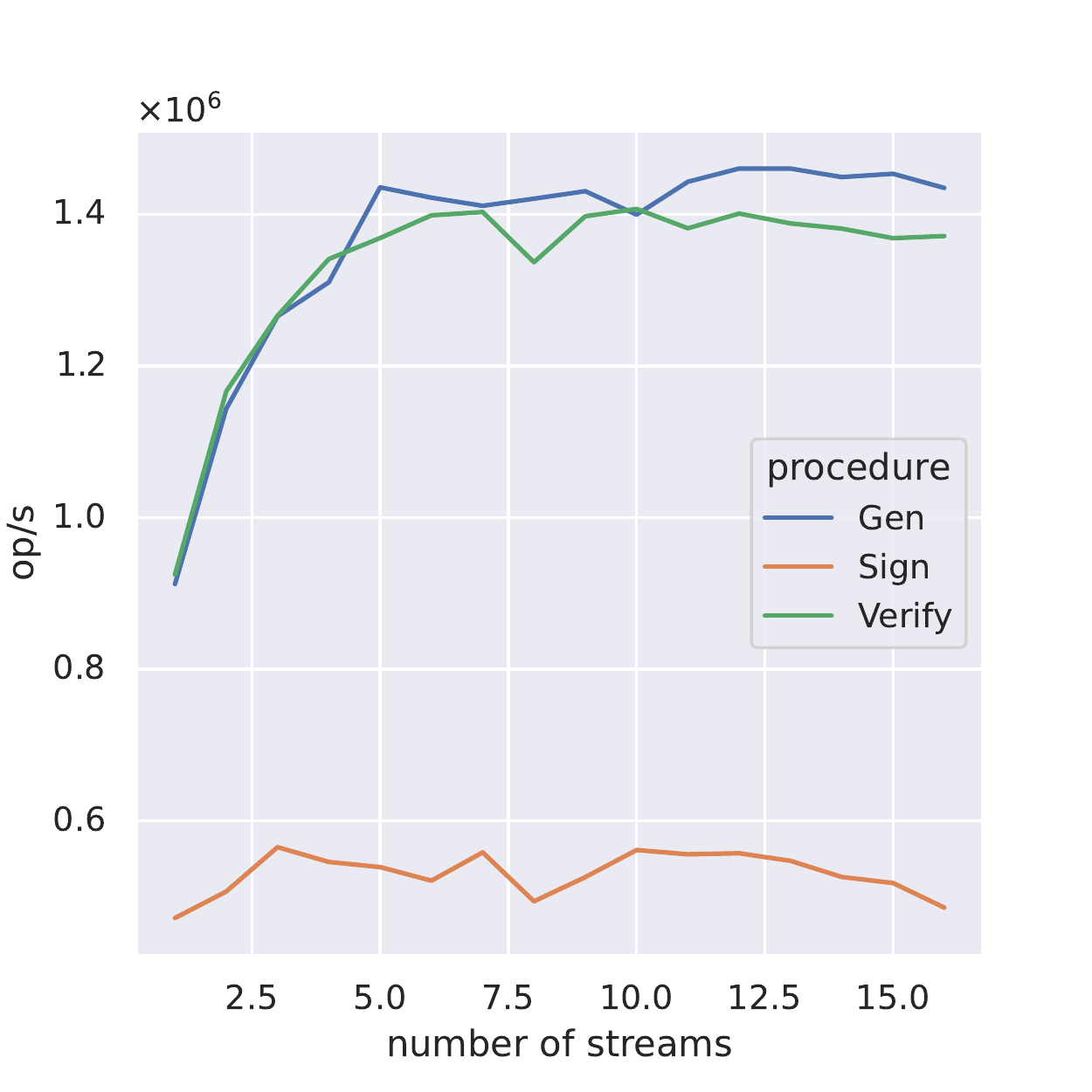}
         \caption{}
         \label{fig:res-stream}
     \end{subfigure}
     \caption{Sensitivity to the (a) batch sizes and (b) number of streams under fixed concurrent processing sizes.}
	\label{fig:res-batch-stream}
\end{figure}

We conduct a sensitivity study of our implementation under various execution settings, using \dd2 as a representative example. Three parameters are involved: the number of batched tasks on the host side ($\Phi$), the number of concurrent processing tasks on the device side ($\Psi$), and the number of launched CUDA streams.
First, we examine the performance of the $\mathsf{Sign}$ procedure under different processing sizes, by setting $\Psi$ in the range $[1000, 10000]$. We present the throughput results for three batch sizes (1000, 5000, and 10000) in Figure \ref{fig:res-threshold}. The figure reveals that the performance exhibits a periodic pattern, with a rise followed by a sustained decline within each period. As the execution size increases, the overall throughput initially improves before declining. For all three batch sizes, the optimal point is reached at around 2000-3000, e.g., for $\Psi = 10000$ the optimal processing size is $\Psi = 2512$. It should be noted that since the randomness of the scheme affects the number of rounds, we can only obtain an optimal interval.
Next, we fix $\Psi$ to 2512 for the $\mathsf{Sign}$ procedure and examine the throughput of all three procedures under different batch sizes ($\Phi$). Based on this, we vary the number of launched streams from 1 to 16 to test the effectiveness of hiding IO latency. Figures \ref{fig:res-batch} and \ref{fig:res-stream} demonstrate that as the batch size increases, the throughput first rises rapidly and then begins to plateau at around 6000. The throughput of the three procedures mostly reaches optimal levels when launching around 10 to 12 streams.

\subsection{Performance on Different GPUs}

\begin{table}[t]
\centering
\setlength{\tabcolsep}{3.0mm}
\caption{Throughput of C and AVX2 implementations on CPU, our implementations on three different GPUs, and related works on GPU \cite{seo2022parallel}, ARM Cortex-A72 \cite{DBLP:journals/tches/BeckerHKYY22}, and FPGA \cite{DBLP:journals/tches/ZhaoZWYZLZYWL22}. The metric is the operations per second (OP/s). The speedups refer to the comparison with CPU reference implementation.}
\label{tab:performance-scheme}
\begin{tabular}{c|c|rr|rrr|rrr}
\hline
\multirow{2}{*}{Level} & \multirow{2}{*}{} & \multicolumn{2}{c|}{CPU}                            & \multicolumn{3}{c|}{Our work}                                                       & \multicolumn{3}{c}{Related Works}                                             \\ \cline{3-10} 
                       &                            & \multicolumn{1}{c}{Ref} & \multicolumn{1}{c|}{AVX2} & \multicolumn{1}{c}{A100} & \multicolumn{1}{c}{V100S} & \multicolumn{1}{c|}{3090 Ti} & \multicolumn{1}{c}{\cite{seo2022parallel}} & \multicolumn{1}{c}{\cite{DBLP:journals/tches/BeckerHKYY22}} & \multicolumn{1}{c}{\cite{DBLP:journals/tches/ZhaoZWYZLZYWL22}} \\ \hline
\multirow{6}{*}{2}     & $\mathsf{Gen}$             & 21,931                  & 67,188                    & 1,400,257                & 120,779                   & 1,418,283                    & 84,993                  & 5,561                    & 23,217                   \\
                       &                            &                         &                           & ($\mathbf{64\times}$)             & (\textbf{6$\times$})               & (\textbf{65$\times$})                 &                         &                          &                          \\
                       & $\mathsf{Sign}$            & 5,182                   & 23,678                    & 574,953                  & 61,727                    & 562,534                      & 33,965                  & 2,310                    & 3,448                    \\
                       &                            &                         &                           & (\textbf{111$\times$})            & (\textbf{12$\times$})              & (\textbf{109$\times$})                &                         &                          &                          \\
                       & $\mathsf{Verify}$          & 20,524                  & 67,084                    & 1,408,703                & 128,373                   & 1,484,420                    & 67,738                  & 5,498                    & 21,904                   \\
                       &                            &                         &                           & (\textbf{69$\times$})             & (\textbf{6$\times$})               & (\textbf{72$\times$})                 &                         &                          &                          \\ \hline
\multirow{6}{*}{3}     & $\mathsf{Gen}$             & 12,163                  & 41,510                    & 882,569                  & 65,536                    & 769,191                      & 51,099                  & 2,908                    & 16,555                   \\
                       &                            &                         &                           & (\textbf{73$\times$})             & (\textbf{5$\times$})               & (\textbf{63$\times$})                 &                         &                          &                          \\
                       & $\mathsf{Sign}$            & 3,396                   & 15,415                    & 396,360                  & 39,040                    & 372,873                      & 14,875                  & 1,377                    & 2,167                    \\
                       &                            &                         &                           & (\textbf{117$\times$})            & (\textbf{11$\times$})              & (\textbf{110$\times$})                &                         &                          &                          \\
                       & $\mathsf{Verify}$          & 13,421                  & 42,210                    & 982,648                  & 73,135                    & 865,119                      & 44,502                  & 3,352                    & 15,671                   \\
                       &                            &                         &                           & (\textbf{73$\times$})             & (\textbf{5$\times$})               & (\textbf{64$\times$})                 &                         &                          &                          \\ \hline
\multirow{6}{*}{5}     & $\mathsf{Gen}$             & 8,299                   & 26,421                    & 599,349                  & 39,610                    & 463,949                      & 31,800                  & 1,916                    & 11,051                   \\
                       &                            &                         &                           & (\textbf{72$\times$})             & (\textbf{5$\times$})               & (\textbf{56$\times$})                 &                         &                          &                          \\
                       & $\mathsf{Sign}$            & 2,669                   & 12,728                    & 315,900                  & 27,787                    & 271,835                      & 20,396                  & 1,044                    & 1,977                    \\
                       &                            &                         &                           & (\textbf{118$\times$})            & (\textbf{10$\times$})              & (\textbf{102$\times$})                &                         &                          &                          \\
                       & $\mathsf{Verify}$          & 8,061                   & 26,839                    & 637,299                  & 40,990                    & 488,732                      & 27,511                  & 1,961                    & 10,716                   \\
                       &                            &                         &                           & (\textbf{79$\times$})             & (\textbf{5$\times$})               & (\textbf{61$\times$})                 &                         &                          &                          \\ \hline
\end{tabular}
\end{table}

In Table \ref{tab:performance-scheme}, we list the performance of our implementation and the comparisons.
The results for the C and AVX2 implementations are obtained by running the official implementation\footnote{\url{https://github.com/pq-crystals/dilithium}} on our platform.
For the closed-source work \cite{seo2022parallel}, we use the reported results from their paper, which were obtained on a Jetson AGX Xavier GPU.
It is worth noting that in \cite{seo2022parallel}, the authors replaced the time-consuming rejection sampling with a simple data loading based on known positions, which may lead to incompatibility in applications.
The works \cite{DBLP:journals/tches/BeckerHKYY22} and \cite{DBLP:journals/tches/ZhaoZWYZLZYWL22} represent the state-of-the-art Neon-based implementation and FPGA design of \dd.
The results we list come from the original paper.
For our GPU implementation, we report the performance using 10 streams.
In the $\mathsf{Sign}$ procedure, each stream executes 10,000 tasks with a concurrent processing size of 2,512.
Compared to the CPU baseline, our implementation achieves $64\times$-$73\times$ improvement for $\mathsf{Gen}$, $111\times$-$118\times$ improvement for $\mathsf{Sign}$, and $69\times$-$79\times$ improvement for $\mathsf{Verify}$ on the A100 GPU.
Furthermore, our implementation demonstrates over a $20\times$ improvement compared to the AVX2 implementation across all security levels.

\section{Conclusion} \label{sec:con}
In this work, we present a high-throughput GPU implementation of the \dd\ post-quantum digital signature scheme, significantly improving its performance. 
By employing a range of computational and memory optimizations, we have addressed various performance bottlenecks, memory usage, and IO latency issues. 
Our work showcases the potential of utilizing GPU resources to enhance the performance of lattice-based cryptographic schemes, providing valuable insights for future implementations and optimizations. The experimental results highlight our implementation's potential to provide real-time and high-throughput solutions for a wide range of applications in real systems, thereby contributing to the ongoing development and adoption of post-quantum cryptography.

%
%
%
\bibliographystyle{splncs04}
\bibliography{ref}

\end{document}